\documentclass[]{emulateapj}
\usepackage{color}
 
\received{}
\accepted{}
\shortauthors{JOHNSTON ET AL}
\shorttitle{CROSS-CORRELATION CLUSTER LENSING IN THE SDSS II}
\begin{document}

\author{
David E. Johnston,\altaffilmark{1,2}
Erin S. Sheldon,\altaffilmark{3}
Risa H. Wechsler,\altaffilmark{8}
Eduardo Rozo,\altaffilmark{9}
Benjamin P. Koester, \altaffilmark{4,11}
Joshua A. Frieman,\altaffilmark{4,10,11}
Timothy A. McKay,\altaffilmark{5,6,7}
August E. Evrard,\altaffilmark{5,6,7}
Matthew R. Becker \altaffilmark{5}
James Annis, \altaffilmark{10}
}
\altaffiltext{1}{Jet Propulsion Laboratory, 4800 Oak Grove Drive, Pasadena CA, 91109, email davej@astro.caltech.edu}
\altaffiltext{2}{California Institute of Technology, 1200 East California Blvd, Pasadena, CA 91125}
\altaffiltext{3}{Center for Cosmology and Particle Physics,Department of Physics, New York University, 
4 Washington Place, New York, NY, 10003} 
\altaffiltext{4}{Department of Astronomy and Astrophysics, The University of Chicago, 5640 South Ellis Avenue, Chicago, IL 60637}
\altaffiltext{5}{Physics Department, University of Michigan, Ann Arbor, MI 48109}
\altaffiltext{6}{Astronomy Department, University of Michigan, Ann Arbor, MI 48109}
\altaffiltext{7}{Michigan Center for Theoretical Physics, Ann Arbor, MI, 48109}
\altaffiltext{8}{Kavli Institute for Particle Astrophysics and Cosmology, Physics
Department, and Stanford Linear Accelerator Center, Stanford University, 382 Pueblo Mall, Stanford, CA, 94305}
\altaffiltext{9}{CCAPP Fellow, The Ohio State University, Columbus, Ohio 43210}
\altaffiltext{10}{Center for Particle Astrophysics, Fermi National Accelerator Laboratory, P.O. Box 500, Batavia, IL 60510}
\altaffiltext{11}{Kavli Institute for Cosmological Physics, The University of
  Chicago, 5640 South Ellis Avenue, Chicago, IL 60637}

\title{Cross-correlation Weak Lensing of SDSS galaxy Clusters II: Cluster Density Profiles 
and the Mass--Richness Relation}

\begin{abstract}
  We interpret and model the statistical weak lensing measurements around 130,000
  groups and clusters of galaxies in the Sloan Digital Sky Survey presented by \citet{sheldon_etal:07}.
  We present non-parametric inversions of the 2D shear profiles to the mean 3D cluster density and mass 
  profiles in bins of both optical richness and cluster $i$-band luminosity. 
  Since the mean cluster density profile is proportional 
  to the cluster--mass correlation function, the mean profile is 
  spherically symmetric by the assumptions of large-scale homogeneity and isotropy.
  We correct the inferred 3D profiles for systematic effects, including non-linear shear 
  and the fact that cluster halos are not all precisely 
  centered on their brightest galaxies. We also 
  model the measured cluster shear profile as a sum of contributions from the brightest 
  central galaxy, the cluster dark matter halo, and neighboring halos. We infer 
  the relations between mean cluster virial mass and optical richness and luminosity over 
  two orders of magnitude in cluster mass; the virial mass at fixed richness or luminosity 
  is determined with a precision of $\sim 13\%$ including both statistical and systematic 
  errors. We also constrain the halo concentration parameter and halo bias as a function 
  of cluster mass; both are in good agreement with predictions from N-body simulations 
  of LCDM models. The methods employed here will be applicable to deeper, wide-area optical 
  surveys that aim to constrain the nature of the dark energy, such as
  the Dark Energy Survey, the Large Synoptic Survey Telescope and space-based surveys.
\end{abstract}


\section{Introduction}
\label{sec:intro}

Clusters of galaxies are among the most promising probes of cosmology
and of the physics of structure formation. Theoretical calculation
\citep{gott-gunn:collapse,press-schecter:halos} followed by numerical
simulations with ever-increasing resolution
\citep[e.g.][]{nfw:profile,evrard:hubble-volume} have led to a robust,
quantitative framework for the understanding of the non-linear growth,
collapse, and evolution of dark-matter halos. Rich clusters are now
confidently associated with the most massive, collapsed halos.  N-body
simulations predict the abundance of halos
\citep{sheth-tormen:cluster-bias, warren_06}, their density profiles
\citep{nfw:profile}, their concentrations
\citep{bullock:concentration, eke_etal:01, wechsler_etal:02,
  maccio_etal:07, neto:concentration}, and their large-scale
clustering \citep{kaiser:cluster-bias,mo-white:cluster-bias,
  seljak_warren:04, wetzel_07}.  The abundance of dark matter halos is a
strong function of the cosmological parameters, especially $\sigma_8$,
the normalization of the matter power spectrum
\citep{white-ef-frenk:clusters,viana-liddle:clusters,bahcall:sdss-clusters}. Moreover,
the evolution of the cluster abundance with redshift is quite
sensitive to the equation of state of the dark energy
\citep{haiman_etal:01, huterer_turner:01, newman_etal:02,
  levine_etal:02}. An accurate measurement of the cluster abundance
can thus be used to determine cosmological parameters.

However, to exploit clusters as cosmological probes requires knowledge
of the relation between their observable properties and their masses
--- so far, a measurement of the cluster mass-observable relation with
the necessary robustness and precision has been lacking. Various
methods have been employed to detect clusters and to estimate their
masses; each has advantages and disadvantages, and it is likely that
in the future they will be increasingly used in combination.
 
Measurements of X-ray flux and temperature profiles, combined with the
assumption that the X-ray emitting gas is in hydrostatic equilibrium
(HSE) in a spherically symmetric gravitational potential, can be used
to infer cluster mass profiles \citep{reiprich:clusters,nagai_07}.
However, recent XMM-Newton and Chandra data
\citep{markevitch:x-ray-clusters} have shown that a fraction of
clusters have complex luminosity and temperature structure, perhaps
associated with recent merger or AGN activity, calling into question
the spherical HSE assumption in those cases
\citep{evrard:xray-masses}.  In addition, inference of the mass
profile in HSE requires measurement of the radial gas temperature
profile, which in turn requires large numbers of X-ray photons, so
only nearby \citep{sanderson:clusters} or very massive clusters
\citep{allen:large-clusters} are suitable for this treatment.  Future
X-ray observatories such as XEUS and Constellation-X will have greatly
improved sensitivity and will therefore be able to probe lower-mass
clusters with this technique.

The Sunyaev-Zel'dovich (SZ) effect, another gas-based method of
detecting clusters \citep{grego:sz-clusters,carlstrom_02}, has the
advantage of being essentially redshift independent. Theoretically,
the integrated Sunyaev-Zel'dovich flux increment is tightly correlated
with cluster mass \citep{motl_05,nagai_06}, and the slope of the
relation appears to be insensitive to gas dynamics in cluster cores.
Challenges for this technique \citep{hallman_06} include the
identification and removal of contamination by radio point sources
\citep{vale_06}.  Recent SZ measurements \citep{laroque_06} will soon
be supplemented by studies from the APEX-SZ, the Sunyaev-Zel'dovich
Array, and by large surveys with, e.g., the Atacama Cosmology
Telescope and the South Pole Telescope.

Dynamical cluster mass estimates, using the estimated velocity
dispersion of cluster galaxies, are also useful, but they require many
spectroscopic measurements per cluster. The interpretation of
the velocity dispersion as a measure of`the cluster mass also usually
requires assumptions about dynamical equilibrium and about the
distribution of galaxy orbits (velocity anisotropy), although
techniques to bypass these assumptions by simulating cluster galaxy
dynamics directly have also been employed \citep{evrard:mass-vel}.
Dynamical estimates are also subject to uncertainty in the relation
between galaxy and dark matter velocity dispersion, called
velocity bias, which in principle requires inclusion of gas dynamics
and stellar feedback to properly simulate.  Recent work indicates that
this effect is small, but depends on the type of galaxy sampled
\citep[e.g.][]{nagai_kravtsov:05, diemand:velbias}.

Gravitational lensing has proven an effective tool in probing the
masses of clusters. Due to the simplicity of the gravitational physics
of lensing, it has become one of the most secure ways of demonstrating
the existence of dark matter \citep{clowe_etal:06}.  Strong lensing,
using multiple images and arcs, can provide precise cluster mass
estimates on small scales \citep{hammer:arcs,kneib:abell2218}.
However, strong lensing only occurs in very massive clusters;
moreover, strong lensing clusters may not be typical of clusters of
their mass, since the existence of arcs requires high central mass
concentrations. Weak lensing has been used to construct projected mass
maps of clusters to larger scales
\citep[e.g.][]{fahlman:cluster-lensing,tyson-fischer:1689,luppino-kaiser:cluster,
  clowe-luppino:cluster,joffre:lensing-3667,dahle:weak-lensing-clusters,
  cypriano:24-xray,bradac:cluster}. However, individual weak lensing
cluster mass estimates inferred from shear measurements are subject to
$\sim 20$\% uncertainties
\citep{metzler1,metzler2,hoekstra_03,deputter_05}, since they are
sensitive to all mass along the line of sight to the source galaxies,
not just that associated with the cluster. Weak lensing mass estimates
are also affected by the ``mass-sheet'' degeneracy
\citep{bradac:mass-sheet}: adding a constant mass sheet to the 2D mass
density does not change the weak lensing shear.

Fortunately, to use clusters to constrain cosmological parameters,
determination of the masses of {\it individual} clusters is
unnecessary, since cosmological predictions of structure formation are
statistical in nature. Cosmological theory robustly predicts the halo
mass function $n(M;z,\theta_i)$, where $\theta_i$ stands for a vector
of cosmological parameters. Astronomical observations measure the
abundance of clusters sorted by some observable property $O$,
$n(O;z)$. To compare theoretical predictions with observations, we
need to measure or constrain the conditional probability distribution,
$P(O|M;z)$, that a dark matter halo of mass $M$ at redshift $z$ will be
observed as a cluster with observable $O$ in a given survey, including
selection effects and biases.  This is the approach employed, e.g., by
\citet{rozo:methods}, who adopt the halo occupation distribution (HOD)
description of this conditional probability distribution and
marginalize over the HOD model parameters to arrive at cosmological
constraints. Alternatively, one could rely on, e.g., hydrodynamic or
semi-analytic galaxy formation models to directly predict
$n(O;z,\theta_i)$, but the theoretical uncertainties --- which are
roughly captured in the HOD model --- are still large.

The method of ``cross-correlation weak lensing'' provides a direct
estimate of the mean mass for clusters with some observable property
$O$ and therefore an important constraint on the probability
distribution $P(O|M;z)$ needed to connect cosmological theory with
cluster observations.  Cross-correlation lensing consists of stacking
the weak lensing signal from a large number of objects, selected by
some property $O$, to measure the average shear profile with high
signal-to-noise.  By combining the signal from many lenses, the error
on the mean shear profile and on the inferred mean mass can in
principle be reduced to the sub-percent level; in that limit,
systematic errors of interpretation start to dominate. Since less
massive objects are more abundant in the Universe, cross-correlation
lensing can be used over a very wide range of lens masses --- from
massive clusters down to galaxies, where it is referred to as
galaxy-galaxy lensing
\citep{tyson:gal-gal,brainerd:gal-gal,fischer:gal-gal-long,sheldon:gmcf,mandelbaum:groups}.
Because the method corresponds to a statistical measurement of the
lens-mass cross-correlation function (see \S \ref{sec:invert}), the
inferred mean masses are insensitive to {\it uncorrelated} mass along
the line of sight to the source galaxies. For cluster-scale lenses,
the mean effects of {\it correlated} mass along the line of sight,
e.g., in neighboring clusters or filaments, are generally negligible
out to scales comparable to the cluster virial radius.
Moreover, their effects can be measured and modeled, as we show in \S
\ref{sec:halofits}. As a result, cross-correlation lensing is
essentially free of the projection effects that plague individual
cluster lens mass estimates.

In \citet{sheldon_etal:07} (hereafter Paper I), 
%Sheldon et al. (Paper I) 
we presented average shear profiles from cross-correlation weak lensing measurements 
around $\sim$ 130,000 clusters of galaxies from the Sloan Digital Sky
Survey \citep[SDSS,][]{york:sdss}. These clusters were selected   
from the maxBCG cluster catalog 
described in \cite{koester:maxbcg-cat}; the maxBCG cluster finding
algorithm, based on the red sequence of early-type cluster galaxies, 
is described in \cite{koester:maxbcg-alg}. 
%The lensing signals are stacked for clusters 
%in 12 bins of galaxy-number richness $N_{200}$ and separately in 16 bins of 
%cluster $i$-band luminosity $L_{200}$ (see \S \ref{sec:richopt}).   

In this paper, we analyze the detected lensing signal presented in
Paper I and model the features seen in the shear profiles. In \S
\ref{sec:meas} we summarize the relevant results from Paper I. In \S
\ref{sec:invert} we apply the non-parametric inversions of
\cite{johnston:inversion} to infer the mean 3D cluster mass density and
aperture mass profiles in bins of optical richness and luminosity (see \S
\ref{sec:richopt}).  These inverted density and mass profiles,
however, {\it cannot} be directly interpreted as profiles of dark
matter halos. In \S \ref{sec:halofits}, we discuss why this is so and
develop a parameterized model which includes the effects of:
displacement of the center of the cluster halo from the brightest
cluster galaxy (BCG); non-linear shear corrections; lensing by the
central BCG; and lensing by neighboring clusters and structures. When
these effects are included, we find that the inferred halo profiles
are well fit by the universal dark matter profiles of Navarro, Frenk
\& White \citep{nfw:profile}.  In the context of this model, we
estimate the average halo virial mass, $M_{200}$, as a function of
cluster galaxy richness and total galaxy luminosity. We infer the mean halo
concentration and halo bias as a function of $M_{200}$ and find them
to be in good agreement with the predictions of N-body simulations for
the standard LCDM cosmology. In \S \ref{sec:dynamical} we compare the
inferred mean halo masses vs. galaxy richness to recent dynamical mass
estimates from measured velocity dispersions for the same cluster
sample \citep{becker:vel}; the two mass estimates agree very well,
with the lensing estimates having smaller errors. We conclude by
discussing some cosmological applications of these results as well as
applications in future optical surveys.

For computing distances and, where needed, the linear power spectrum
of density perturbations, we use a spatially flat cosmological model
with a cosmological constant and cold dark matter (LCDM) with scaled
CDM density $\Omega_m=0.27$, baryon density $\Omega_b=0.045$, scaled
Hubble parameter $h=0.71$ (for the linear power spectrum not distances) 
and primordial spectral index $n_s=0.95$. 
The linear power spectrum amplitude $\sigma_8$ is left free except where specified. 
We employ the linear transfer function of Eisenstein and Hu
\citep{eisenstein-hu:transfer}. This model (with $\sigma_8=0.8$) fits
both the WMAP third-year data \cite{spergel:wmap3} and the SDSS
luminous red galaxy (LRG) clustering data \citep{eisenstein:wiggles}.
All distances in this paper are in \emph{physical} not \emph{comoving} units
of $h^{-1} Mpc$.

\section{Weak lensing shear measurements}
\label{sec:meas}

The methods of measuring the weak lensing signal are described in
detail in Paper I. We briefly summarize some of the important features
here. For any projected mass distribution, the azimuthally averaged tangential shear at
projected radius $R$ from the center of the distribution is given by
$\gamma(R) = \Delta \Sigma(R)/\Sigma_{crit} \equiv
[\overline{\Sigma}(< R) - \overline{\Sigma}(R)]/\Sigma_{crit}$, where
$\Sigma(R)$ is the 2D projected mass density at radius $R$,
$\overline{\Sigma}(< R)$ is the average of $\Sigma$ inside a disk of
radius $R$, $\overline{\Sigma}(R)$ is the azimuthal average of
$\Sigma(R)$ in a thin annulus of radius $R$, and the critical density
for strong lensing is given by $\Sigma_{crit} \equiv c^2/(4 \pi G)
~D_S/(D_L D_{LS})$, with $D_S, D_L, D_{LS}$ the angular diameter
distances from the observer to the source, to the lens, and between
the lens and source, respectively. These distances are
cosmology-dependent functions of redshift. Paper I presents average
profiles of $\Delta \Sigma(R)$ for maxBCG clusters binned by cluster galaxy number, $N_{200}$,
and by optical luminosity $L_{200}$. For these measurements, the radius $R$ is
defined with respect to the position of the BCG; see \S
\ref{sec:miscenter} for further discussion of this point.

\subsection{Richness and Luminosity measures $N_{200}$ and $L_{200}$}
\label{sec:richopt}

Although the richness and luminosity measures $N_{200}$ and $L_{200}$
are discussed in detail in Paper I, here we emphasize some of their
important features to avoid possible confusion. $N_{200}$ and
$L_{200}$ are the galaxy number and total $i$-band luminosity measured
within a projected radius we call $r_{200}^{gals}$, in both
cases counting only red-sequence galaxies with luminosities larger
than $0.4L_*$ and satisfying other selection criteria (see
\citealt{koester:maxbcg-alg} for details).  This radius is {\it not}
by definition, equivalent to the $r_{200}$ defined by the mass (Eqn. \ref{eq:virMr}),
which can in principle be measured directly from lensing, since
$r_{200}$ is not known prior to performing the weak lensing
analysis. Instead, $r_{200}^{gals}$ is determined by first
measuring the number of galaxies, $N_{gal}$, within a fixed 1 $h^{-1}$
Mpc aperture and calculating $r_{200}^{gals} = 0.156
~N_{gal}^{0.6} ~h^{-1}$ Mpc, as discussed in \cite{hansen:r200}.
Nevertheless, we find that $r_{200}^{gals}$ is in fact a good approximation
to $r_{200}$ as determined in this paper from the lensing data to within about 5\%.
%Thus, $N_{200}$ and $L_{200}$ are \emph{not} the total number of galaxies and luminosity within the radius $r_{200}$. 
The mass-to-light ratio as a function of radius will be presented in Paper III of this series (Sheldon et al. 2007). 
Note that $N_{200}$ is dimensionless, and $L_{200}$ has units of $10^{10} h^{-2} L_{\sun}$.

For the purpose of lensing measurement, the catalog is subdivided into 12 $N_{200}$ richness 
bins and 16 $L_{200}$ richness bins. The richness boundaries for each richness measure as well as the number of 
clusters per bin are displayed in Tables \ref{tab:rich-n200} and \ref{tab:rich-l200}.

\begin{deluxetable}{ccc}
\tablecaption{12 $N_{200}$ bins}
\tablewidth{0pt}
\tablehead{
\colhead{Bin number} &
\colhead{$N_{200}$} &
\colhead{Number of clusters per bin}
}
\startdata
1 & 3 & 58788 \\ 
2 & 4 & 27083 \\
3 & 5 & 14925 \\
4 & 6 & 8744 \\
5 & 7 & 5630 \\
6 & 8 & 3858 \\
7 & 9-11 & 6196 \\
8 & 12-17 & 4427 \\
9 & 18-25 & 1711 \\
10 & 26-40 &  787 \\
11 & 41-70 &  272 \\
12 & 71-220 & 47
\enddata
\tablecomments{The catalog is divided into 12 $N_{200}$ richness bins. This table shows
the boundaries of $N_{200}$ values and the number of clusters for each bin.
}
\label{tab:rich-n200}
\end{deluxetable}

\begin{deluxetable}{ccc}
\tablecaption{16 $L_{200}$ bins}
\tablewidth{0pt}
\tablehead{
\colhead{Bin number} &
\colhead{$L_{200} (10^{10} h^{-2} L_{\sun})$} &
\colhead{Number of clusters per bin}
}
\startdata
1 & 5 - 6.24 & 19618 \\ 
2 & 6.24 - 7.8 & 18597 \\
3 & 7.8 - 9.74 & 16042 \\
4 & 9.74 - 12.2 & 12269 \\
5 & 12.2- 15.2 & 9010 \\
6 & 15.2 - 19.0 & 6152 \\
7 & 19.0 - 23.7 & 4164 \\
8 & 23.7 - 29.6 & 2666 \\
9 & 29.6 - 36.9 & 1703 \\
10 & 36.9 - 46.1 &  1042 \\
11 & 46.1 - 57.6 &  638 \\
12 & 57.6 - 71.9 & 344 \\
13 & 71.9 - 89.8 & 210 \\
14 & 89.8 - 112.1 & 108 \\
15 & 112.1 - 140 & 49 \\
16 & 140 - 450 & 46 \\
\enddata
\tablecomments{
The catalog is also divided into 16 $L_{200}$ richness bins. This table shows
the boundaries of $L_{200}$ values and the number of clusters for each bin.
}
\label{tab:rich-l200}
\end{deluxetable}

\begin{figure*}
\epsscale{1.1}
\plottwo{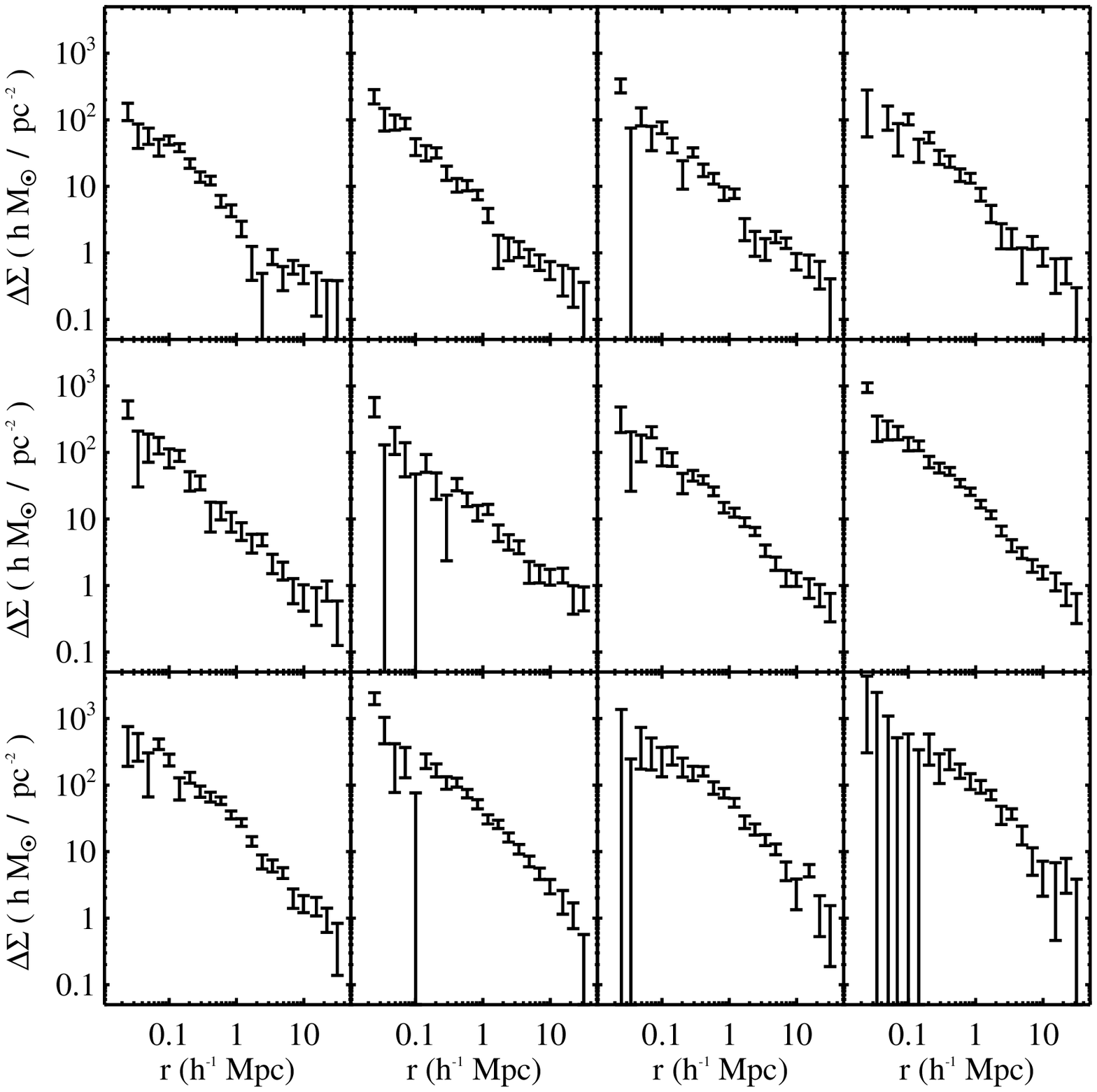}{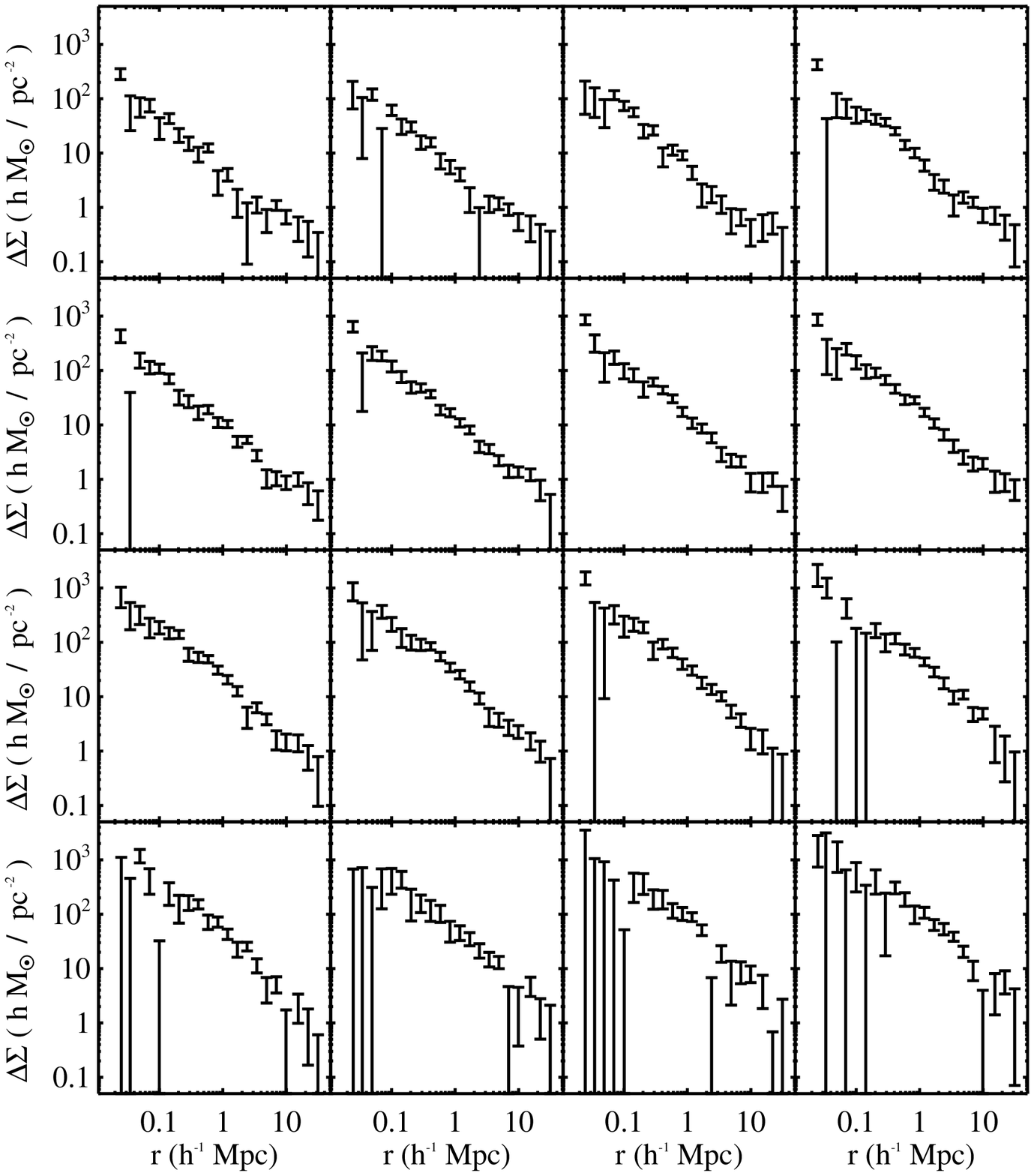}
\caption{{\bf Left:} Weak lensing profiles $\Delta\Sigma(R)$ for 12 bins 
of optical richness, $N_{200}$.
{\bf Right:}
$\Delta\Sigma(R)$ for 16 $i$-band luminosity bins, $L_{200}$.
}
\label{fig:ds}
\end{figure*}

%\section{Method for Inverting Cluster Profiles}
\section{Inverting Cluster Profiles}
\label{sec:invert}
\subsection{Inversion Method}
%Review of the inversion methods}

The methods used to invert the lensing $\Delta\Sigma(R)$ profiles to 3D
density and mass profiles are discussed in detail in
\cite{johnston:inversion} and were first used by \cite{sheldon:gmcf} 
to obtain the galaxy-mass correlation
function from galaxy-galaxy lensing measurements. 
Here, we provide a brief overview of the methods.

The mean excess 3D density profile $\Delta \rho(r)$ around a set of clusters with a given  
observable $O$ (e.g., richness or luminosity) is best thought of in 
terms of the cluster--mass two-point correlation function, $\xi_{cm}$, since  
$\Delta \rho(r) = \bar{\rho}~\xi_{cm}(r)$, where
$\bar{\rho}$ is the mean density of the Universe. By the assumptions of 
spatial homogeneity and isotropy, $\xi_{cm}$ depends only on the 
magnitude of the separation, $r$, not on direction. As a consequence, 
the mean density profile $\Delta \rho(r)$ should be very nearly 
spherically symmetric. Note that this is a purely statistical 
statement: we do {\it not} assume that individual cluster density 
profiles are spherically symmetric. 
The spherical symmetry of the average density 
profile enables the inversion of the stacked 
lensing signal $\Delta\Sigma(R)$ to the 3D
density $\Delta\rho(R)$ and the aperture mass $M(R)$. By contrast, 
weak lensing measurements of {\it individual} clusters can only be 
used to reconstruct the projected 2D mass density,
$\Sigma(\vec{x})$, since lensing is produced by all of the
mass projected along the line of sight.

The mean 3D density profile is obtained as an integral of the derivative of the
shear profile $\Delta\Sigma(R)$ through a purely geometric
relation,

\begin{equation}
\Delta\rho(r) = \frac{1}{\pi} \int_r^{\infty}dR~ \frac{-\Sigma^{\prime}(R)}{\sqrt{R^2-r^2}} ~,
\label{eq:delta-rho}
\end{equation}

\noindent where a prime denotes a derivative with respect to $R$. 
The lensing data $\Delta \Sigma$ enters here since it can be shown that
\begin{equation}
-\Sigma^{\prime}(R) = \Delta\Sigma^{\prime}(R) + 2\Delta\Sigma(R)/R~.
\end{equation}

\noindent The 3D mass profile is given in terms of $\Delta\Sigma(R)$ and $\Delta\rho(R)$
as

\begin{eqnarray}
M(R) = \pi R^2 \Delta\Sigma(R) + 2 \pi \int_R^{\infty} dr~r~\Delta\rho(r) \times \nonumber
\\
\left[ \frac{R^2}{\sqrt{r^2-R^2}} - 2 \left( r - \sqrt{r^2-R^2} \right) \right].
\label{eq:MassInv}
\end{eqnarray}

\noindent In practice, 
these integrals must be truncated at some maximum radius, $R_{max}$, the largest scale
at which one has lensing data ($30 h^{-1}$ Mpc for our data).
The uncertainty from this truncation is
related to the mass-sheet degeneracy. Due to the steepness of the
cluster profiles we infer in this paper, this truncation creates only a few percent
uncertainty in the last few radial bins of both density or mass 
and virtually none in bins at smaller radii.  Complete details of the procedure are given in
\cite{johnston:inversion}.

\subsection{3D Density and Mass Profiles}

\begin{figure*}
\epsscale{1.1}
\plottwo{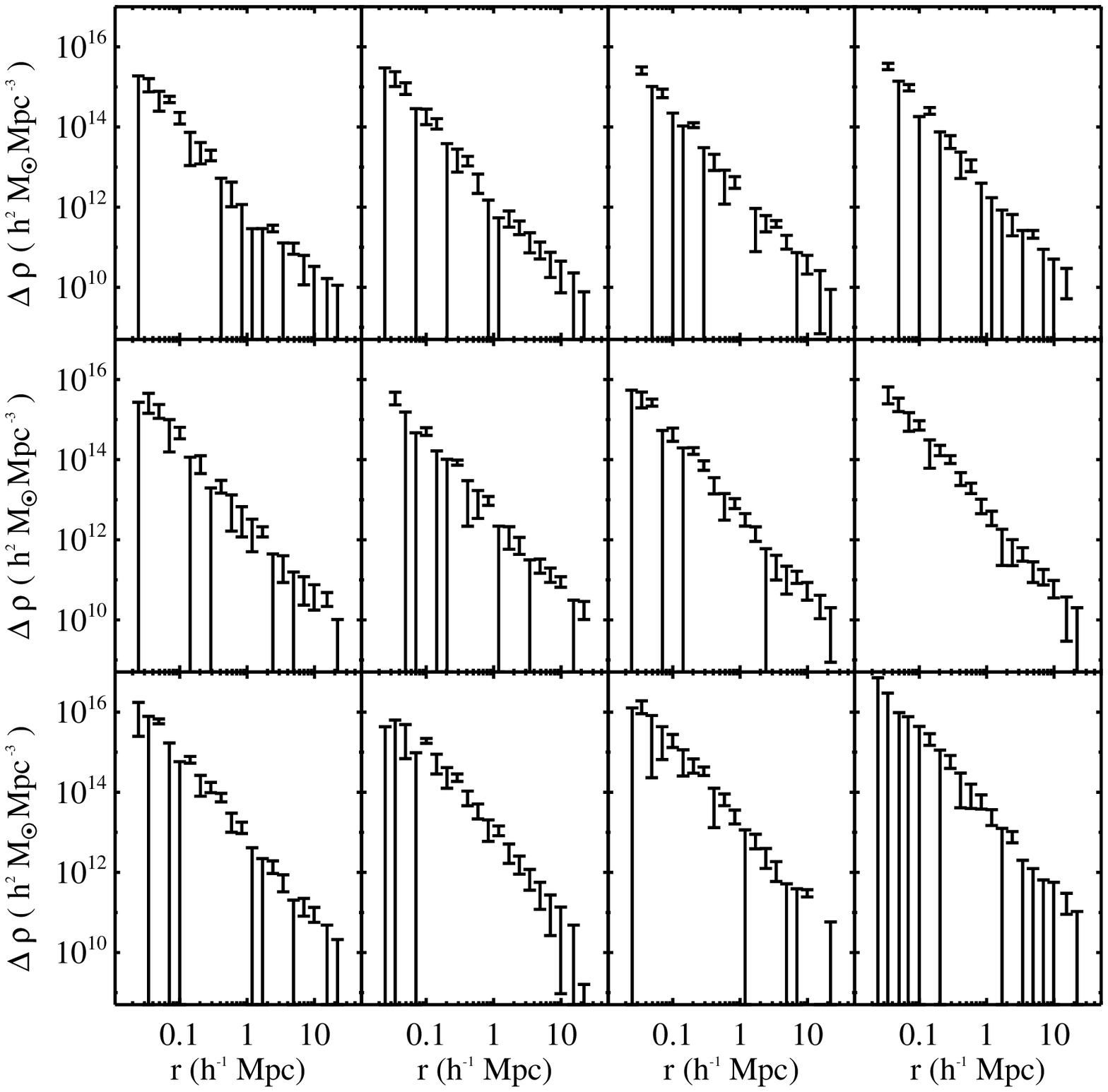}{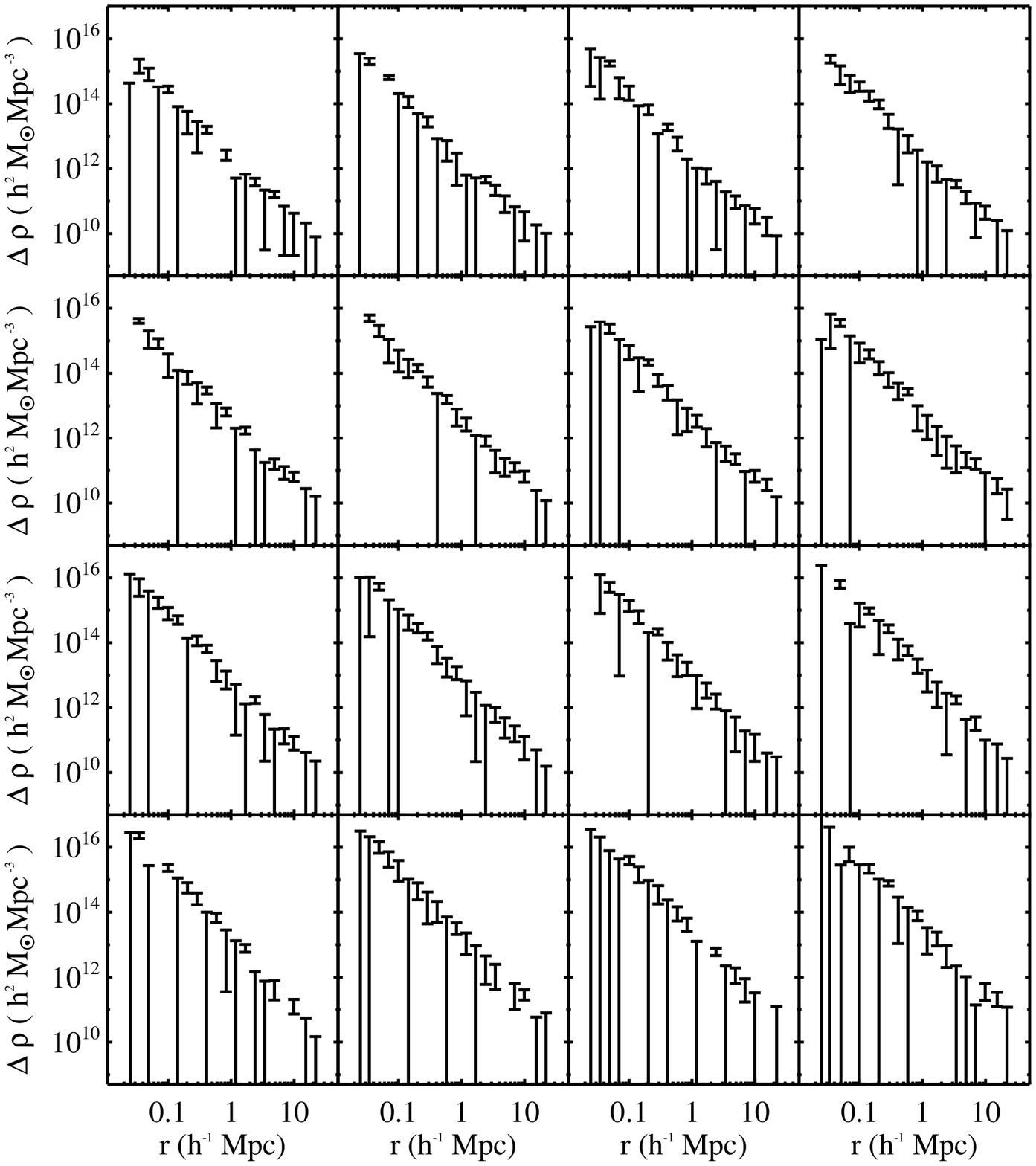}
\caption{{\bf Left:} Inverted mean density profiles, $\Delta\rho(r)$, for the 12 $N_{200}$
  richness bins shown in Fig. \ref{fig:ds}. 
% \label{fig:drho-ngal}}
%\end{figure}
%\begin{figure}
%epsscale{1.0}
%\plotone
%\caption{
{\bf Right:}
Inverted $\Delta\rho(r)$ profiles for the 16 $L_{200}$ richness bins shown in Fig. 
\ref{fig:ds}.
}
%\label{fig:drho-lum}}
\label{fig:drho}
\end{figure*}

The inverted 3D density profiles for each of the 12 $N_{200}$
richness and 16 $L_{200}$ luminosity bins 
are presented in Figure \ref{fig:drho}. These
profiles are noisier than the shear profiles, since they involve 
derivatives of noisy data. The differentiation in Eqn. \ref{eq:delta-rho}
also leads to anti-correlations between neighboring radial bins of $\Delta \rho(r)$. 
%Although these profiles all appear to be power-laws to the eye,
%there are detectable deviations. These are best seen in the mass
%profile plots.

\begin{figure*}
\epsscale{1.1}
\plottwo{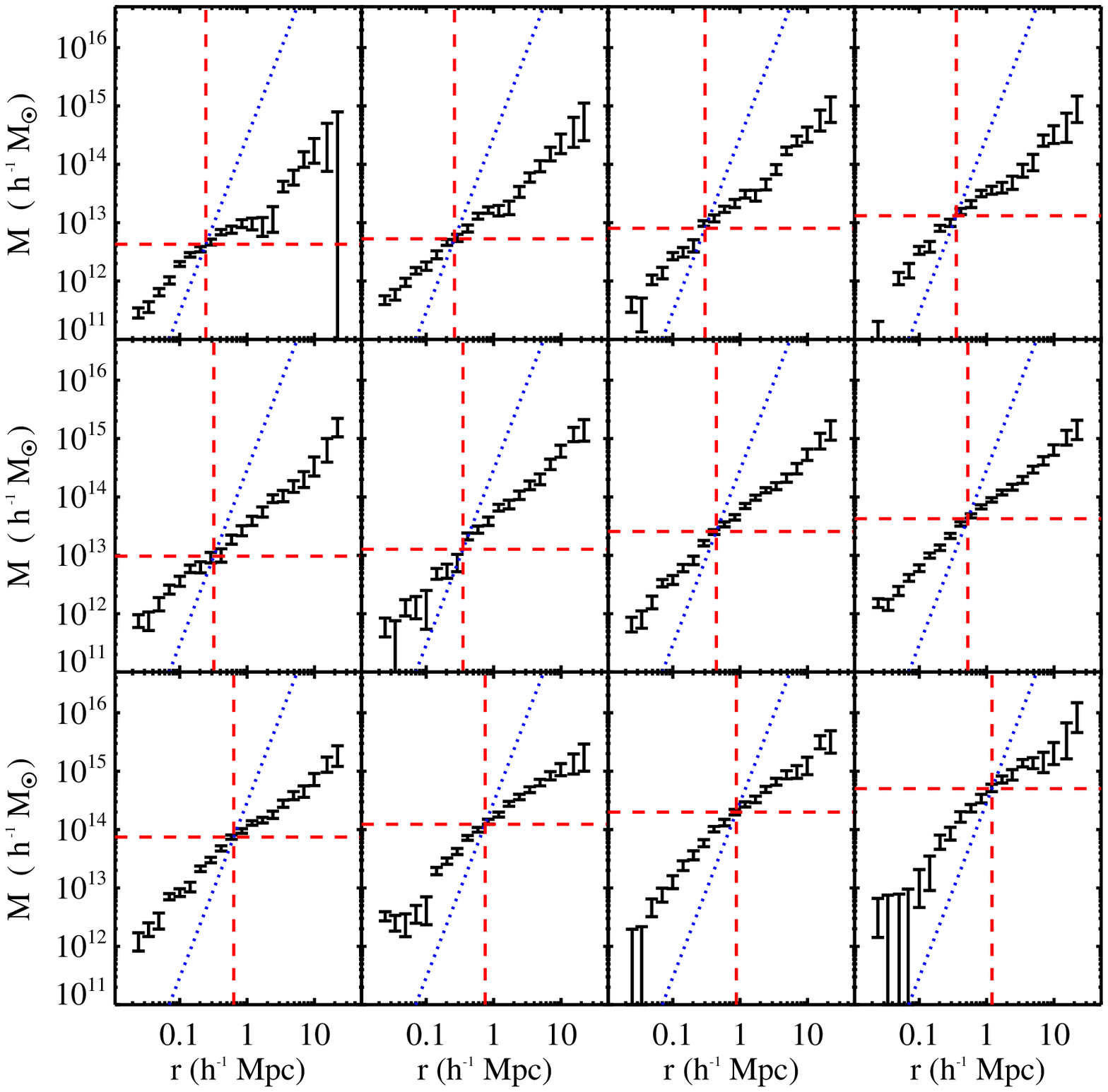}{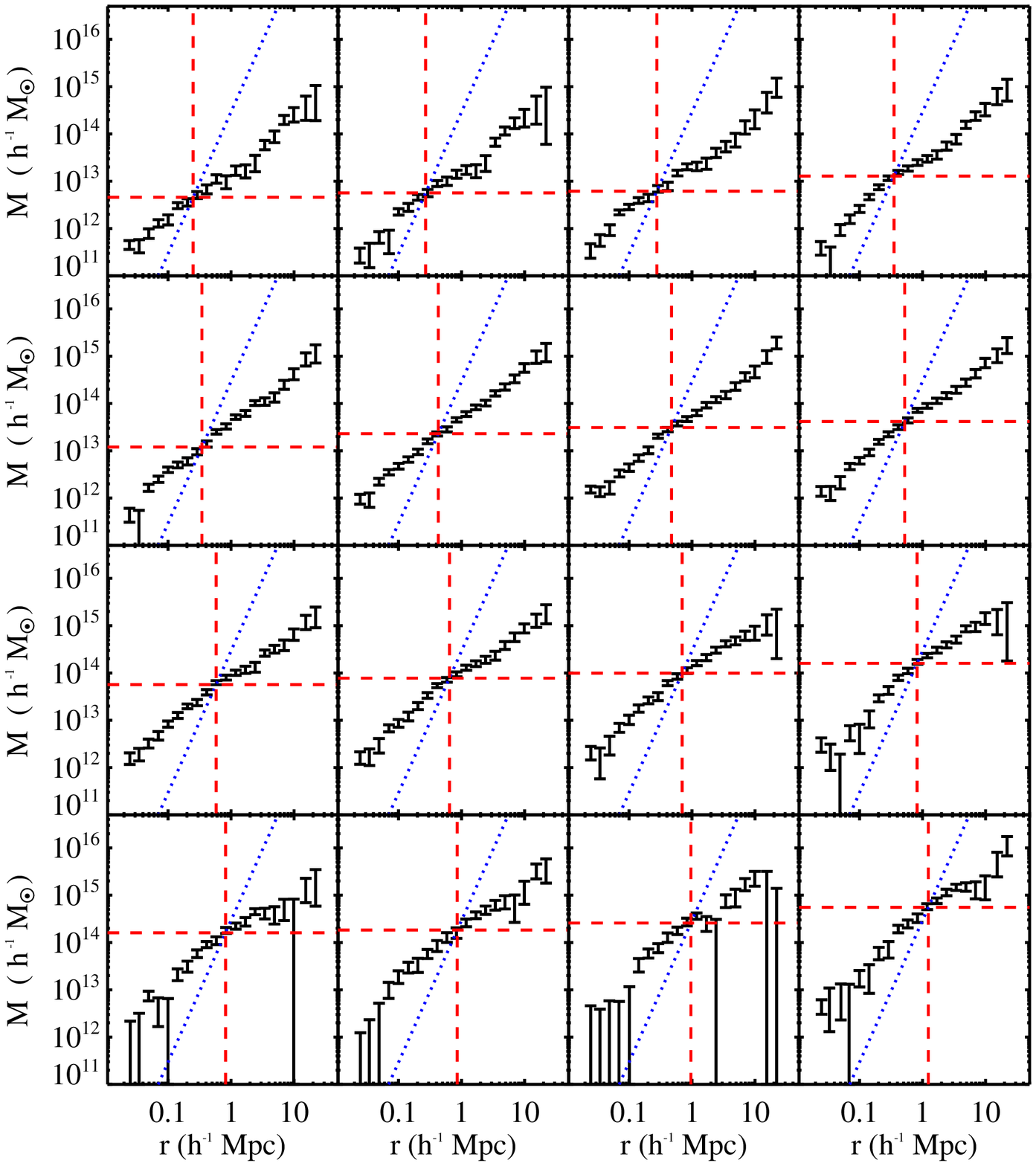}
\caption{
{\bf Left:} 
Inverted 3D aperture mass profiles, $M(r)$, for the 12 $N_{200}$ richness bins. 
The dotted blue diagonal line in each panel denotes
$200~\rho_{c}~4/3 ~\pi~r^3$ (see Eqn. \ref{eq:virMr});
this crosses the mass 
profile at $r_{200}$ and $M_{200}$, which are 
indicated with the dashed red vertical and horizontal lines.
% \label{fig:mass-ngal}}
%\end{figure}
%\begin{figure}
%\epsscale{1.0}
%\plotone
%\caption{
{\bf Right:} 
Inverted 3D aperture mass profiles, $M(r)$, for the 16 $L_{200}$ richness bins.
}
%\label{fig:mass-lum}}
\label{fig:mass}
\end{figure*}

Figure \ref{fig:mass} shows the inverted mean aperture mass profiles,
$M(r)$, for the same richness and luminosity bins as above. Since the mass profile is
an integral of the density profile, it is smoother than the latter, and 
neighboring bins of $M(r)$ 
are statistically correlated. This allows one to better see
the deviations from power-law behavior that one expects from the halo
model (see \S \ref{sec:halofits}).  

\subsection{Direct Measurements of $r_{200}$ and $M_{200}$}
\label{sec:r200M200}
The radius $r_{200}$ is defined, herein, as the radius within which the
average density is 200 times the critical density $\rho_c$. This 
defines the corresponding mass scale, 

\begin{equation}
M_{200} \equiv M(r_{200}) = 200 ~\rho_c(z)~ \frac{4}{3}~ \pi~ r_{200}^3~,
\label{eq:virMr}
\end{equation}

\noindent where $\rho_c(z) = 3 H^2(z)/(8\pi G)$ is the critical density 
at epoch $z$, and the Hubble parameter satisfies 
$H^2(z) = H_0^2 \left[\Omega_m (1+z)^3
  + (1-\Omega_m) \right]$ for a flat LCDM Universe. Throughout this paper we
use $z=0.25$, the mean cluster redshift for our sample, and $\Omega_m=0.27$
to compute $\rho_c(z)$. For these choices, the conversion
between $M_{200}$ and $r_{200}$ is 
$M_{200} = 2.923 \times 10^{14}~ h^{-1} M_{\sun} ~\left(
  r_{200}/h^{-1} {\rm Mpc} \right)^3$.

Using the inverted mass profiles shown in Fig. \ref{fig:mass}, we can
determine $r_{200}$ and $M_{200}$ in a model independent way, by
simply measuring where the curve $200 ~\rho_c(z)~4/3~ \pi~ r^3$
crosses the mass profile. This procedure, which is illustrated in
Fig. \ref{fig:mass}, requires one to interpolate the data between the
two radii closest to the crossing point. This interpolation can in
principle be ill-defined if the data are noisy and the profiles
non-monotonic, but that never occurs for any of our profiles. We have
experimented with a few different ways of interpolating. One can use
the unique power law defined by the two neighboring data points or fit
a power law to a four-point neighborhood of the crossing. We find that
these methods give essentially identical answers; since the four-point
method yields slightly lower scatter in the mass--richness relation,
we use that method.

While this procedure for inferring $r_{200}$ and $M_{200}$ has the
advantage of being model independent, the results {\it cannot} be
interpreted as the virial radii and masses of the corresponding dark
halos. The primary reason is that BCGs, which we use to define the
center of each cluster for the lensing measurements, are not always
positioned at the center of mass of the underlying dark matter halo. This
fact, which we observe in our simulations, is not surprising: for this
analysis, clusters are the objects identified by the maxBCG algorithm,
while dark matter halos are theoretical constructs --- the two are not
in precise one-to-one correspondence
\citep{cohn:red-sequence,rozo:methods}. The model-independent profiles
of Fig. \ref{fig:mass}, and the corresponding values of $r_{200}$ and
$M_{200}$, are the ``true'' mass profiles of clusters centered on
their BCGs. However, to estimate dark matter halo profiles and masses,
we must adopt a model to describe the data, which we do in the next
section. When we do so, we find that the inferred dark matter halo masses
are about 50\% higher than the model-independent cluster
masses. We use the results of those model fits to constrain the halo
mass -- richness relations and other scaling relations.

We will distinguish these two types of masses by referring to the
parametric halo masses as $M_{200}$ and non-parametric cluster masses
as $M_{200}^{cl}$. For completeness, we present these cluster masses
in Tables \ref{tab:mass-rich-direct-ngal} and
\ref{tab:mass-rich-direct-lum} but we will not use them elsewhere in
this work. In another publication, Paper III of this series 
on cluster mass-to-light ratios (Sheldon et al.), we will refer to these non-parametric
$M_{200}^{cl}$ masses.

\begin{deluxetable}{ccc}
\tablecaption{Direct cluster $M_{200}^{cl}$-richness calibration: $N_{200}$ Bins}
\tablewidth{0pt}
\tablehead{
\colhead{$\left<N_{200}\right>$} &
\colhead{$M_{200}^{cl}$ ( $10^{12}h^{-1}M_{\sun}$ )} &
\colhead{$r_{200}^{cl}$ ($h^{-1}$ Mpc)}
}
\startdata
      3.00 &       4.26 $\pm$       0.45 &       0.24 $\pm$       0.01 \\
      4.00 &       5.29 $\pm$       0.65 &       0.26 $\pm$       0.01 \\
      5.00 &       8.01 $\pm$       1.34 &       0.30 $\pm$       0.02 \\
      6.00 &      13.15 $\pm$       1.65 &       0.36 $\pm$       0.01 \\
      7.00 &       9.66 $\pm$       2.28 &       0.32 $\pm$       0.03 \\
      8.00 &      12.71 $\pm$       3.36 &       0.35 $\pm$       0.03 \\
      9.82 &      25.53 $\pm$       2.86 &       0.44 $\pm$       0.02 \\
     13.91 &      42.31 $\pm$       3.42 &       0.53 $\pm$       0.01 \\
     20.78 &      74.45 $\pm$       7.46 &       0.63 $\pm$       0.02 \\
     31.09 &     123.22 $\pm$      11.28 &       0.75 $\pm$       0.02 \\
     50.27 &     199.26 $\pm$      24.81 &       0.88 $\pm$       0.04 \\
     92.18 &     502.87 $\pm$      87.61 &       1.20 $\pm$       0.07 \\
\enddata
\tablecomments{
The $M_{200}^{cl}$ -- richness relation for the $N_{200}$ richness bins. This estimate of $M_{200}$ which we call $M_{200}^{cl}$ is meant to represent the $M_{200}$ of the clusters as opposed to the dark matter halos. It is estimated non-parametrically by determining where the 3D mass profile $M(r)$ cross the line determined by $4/3 \pi r^3 ~200~\rho_{crit}(z)$. These masses differ from the parametric masses that include cluster miscentering and other effects.
}
\label{tab:mass-rich-direct-ngal}
\end{deluxetable}

\begin{deluxetable}{ccc}
\tablecaption{Direct cluster $M_{200}^{cl}$-richness calibration: $L_{200}$ Bins}
\tablewidth{0pt}
\tablehead{
\colhead{$\left<L_{200}\right>$} &
\colhead{$M_{200}^{cl}$ ( $10^{12}h^{-1}M_{\sun}$ )} &
\colhead{$r_{200}^{cl}$ ($h^{-1}$ Mpc)}
}
\startdata
      5.59 &       4.59 $\pm$       0.77 &       0.25 $\pm$       0.01 \\
      6.97 &       5.68 $\pm$       0.81 &       0.27 $\pm$       0.01 \\
      8.69 &       6.16 $\pm$       0.96 &       0.28 $\pm$       0.01 \\
     10.84 &      12.86 $\pm$       1.42 &       0.35 $\pm$       0.01 \\
     13.53 &      11.98 $\pm$       1.80 &       0.34 $\pm$       0.02 \\
     16.89 &      22.92 $\pm$       2.89 &       0.43 $\pm$       0.02 \\
     21.06 &      30.94 $\pm$       3.60 &       0.47 $\pm$       0.02 \\
     26.31 &      41.36 $\pm$       4.51 &       0.52 $\pm$       0.02 \\
     32.89 &      56.90 $\pm$       7.80 &       0.58 $\pm$       0.03 \\
     40.95 &      77.67 $\pm$       9.78 &       0.64 $\pm$       0.03 \\
     51.19 &      99.05 $\pm$      13.49 &       0.70 $\pm$       0.03 \\
     64.08 &     160.65 $\pm$      22.19 &       0.82 $\pm$       0.04 \\
     79.89 &     160.16 $\pm$      30.13 &       0.82 $\pm$       0.05 \\
     98.69 &     182.81 $\pm$      35.58 &       0.86 $\pm$       0.06 \\
    124.59 &     258.49 $\pm$      53.30 &       0.96 $\pm$       0.07 \\
    184.65 &     553.76 $\pm$      93.41 &       1.24 $\pm$       0.07 \\
\enddata
\tablecomments{
The $M_{200}^{cl}$ -- richness relation for the $L_{200}$ richness bins.
}
\label{tab:mass-rich-direct-lum}
\end{deluxetable}

\section{Halo model fits to Lensing Profiles}
\label{sec:halofits}

To proceed, we construct a physical model of the average mass density in clusters 
that comprises three components: the central BCG, the cluster-scale dark matter halo 
in which it sits, and neighboring mass concentrations. We will also consider non-linear shear.
We treat these in turn.

\subsection{The BCG}
\label{sec:bcg}

Since every maxBCG cluster, by design, is centered on a bright galaxy, we should
allow for a contribution to the mass from the baryons (mainly stars)
and from the dark matter sub-halo of the BCG (assuming the latter is
not modeled by the central cusp of the cluster-scale halo).
\cite{gavazzi:slacs} find that a central baryonic component is
required to fit both the strong and weak lensing profiles of
early-type galaxies in the SLACS survey. Although this contribution
could be modeled in a number of ways, e.g., by using a de Vaucouleurs
profile, its effects are only significant on very small scales, and
its form is not well constrained by our data. Therefore, we simply
model this contribution as a central point mass, $M_0$, with lensing signal
$\Delta\Sigma = M_0/(\pi R^2)$, where $M_0$ is a model parameter to be
fit.

\subsection{The cluster dark matter halo}
\label{sec:halo-model}
Out to radii of a few Mpc, the density profiles appear to be dominated by 
the cluster-scale dark matter halos. N-body simulations of structure 
formation with cold dark matter indicate 
that halos are reasonably well modeled by the universal (NFW) profiles of 
\citet{nfw:profile}, 

\begin{equation}
\rho_{NFW}(r) = \frac{\delta~\rho_c(z)}{(r/r_s)(1+r/r_s)^2}~.
\end{equation}

\noindent This form contains two free parameters, a scale radius 
$r_s$ and an amplitude $\delta$; $\rho_c(z)$ is the
critical density at redshift $z$.  At $r \sim r_s$, 
the logarithmic slope of the NFW profile changes between the 
asymptotic values of $-1$ at small scales ($r \ll r_s$) and $-3$ at large 
scales ($r \gg r_s$). The parameters $\delta$ and $r_s$ are usually traded for 
a description in terms of 
$r_{200}$ (or equivalently $M_{200}$) and $c_{200}$. As above, 
$r_{200}$ is the radius within which the mean density is 200 times 
the critical density and for which the enclosed mass is 
$M(r_{200}) = 200\rho_c(z) (4/3) \pi~
r_{200}^3$, while $c_{200} \equiv r_{200}/r_s$ is the concentration.  
The amplitude $\delta$ can be expressed in terms of
$c_{200}$ as

\begin{equation}
\delta=\frac{200}{3}\frac{c_{200}^3}{\ln(1+c_{200})-c_{200}/(1+c_{200})}~.
\end{equation}

\noindent Analytic expressions for the shear 
profile $\Delta\Sigma(R;c_{200},M_{200})$ of NFW halos can be
found in, e.g., \cite{wright:nfw}. Various other 
definitions of the virial radius have been used in the literature, e.g., 
the radius within which the mean density is 
$180~\bar{\rho}(z)$ instead of $200 \rho_c(z)$. We discuss the conversion
among these different systems in the Appendix.

\subsection{Miscentering of the BCG and the Halo} 
\label{sec:miscenter} 
 
For the lensing measurements, the center of each cluster ($R=0$) is
defined to be the position of the BCG identified by the
cluster-finding algorithm.  As noted in \S \ref{sec:r200M200}, some
fraction of the BCGs may be offset from the centers of the
corresponding dark matter halos. Such ``miscentering'' changes the
observed tangential shear profile. If the 2D offset in the lens plane
is $R_s$ then the azimuthally averaged $\Sigma(R)$ profile is given by
the convolution
\begin{equation} 
  \Sigma(R | R_s) = \frac{1}{2 \pi} \int_0^{2 \pi} d\theta \Sigma  
  \left( \sqrt{R^2 + R_s^2 + 2 R~R_s\cos(\theta)} \right)  
\label{eq:conv} 
\end{equation} 
\citep{yang:clust-lens1}.  
 
To make progress, we need to know something about the distribution of
offsets, $P(R_s)$.  In order to estimate this, we employ N-body
simulation-based mock galaxy catalogs that have been constrained to
have realistic luminosities, colors, clustering properties, and
cluster populations.  These catalogs, which have been used in previous
maxBCG studies
\citep{koester:maxbcg-alg,rozo:methods,rozo:mass-function1}, populate a
dark matter simulation with galaxies using the ADDGALS technique
(Wechsler et al 2007).  The catalog is based on the light-cone from
the Hubble Volume simulation \citep{evrard:hubble-volume}, and extends from $0 < z
< 0.34$.  Galaxies are assigned directly to dark matter particles in
the simulation, with a luminosity-dependent bias scheme that is tuned
to match local clustering data.  The galaxy luminosities are first
assigned in the $^{0.1}r$-band, drawn from the luminosity function of
\citet{BlantonLum03}.  The luminosity function is assumed to evolve
with $Q=1.3$ magnitudes per unit redshift.  We first constrain the
relationship between galaxy luminosity and Lagrangian matter densities
on a scale of $\sim M*$, using the luminosity-dependent two-point
clustering of SDSS galaxies \citep{Zehavi05}.  For each galaxy, a dark
matter particle is then chosen on the basis of this density with some
$P(\delta | M_r)$.  Each mock galaxy is then assigned to a real SDSS
galaxy that has approximately the same luminosity and local galaxy
density, measured here as the distance to the fifth nearest neighbor.
The color for each mock is then given by the SED of this matched
galaxy transformed to the appropriate redshift.  Because BCGs are now
known to be distinct from the general galaxy population, BCG
properties are further tuned to match the luminosities and colors of
observed BCGs; in addition a BCG is placed at the center of each dark
matter halo.  This procedure produces a catalog which matches several
statistics of the observed SDSS population, including the location,
width and evolution of the ridgeline, which makes it ideal for testing
the maxBCG algorithm.  In this work, we use five galaxy realizations
that have been run using the same underlying dark matter simulation;
to improve our statistics, we merge all five mock catalogs into one.

The maxBCG algorithm is then used to identify clusters in the mock
catalogs, and the resulting BCG positions can be compared to the
centers of mass of the dark matter halos in the input N-body
simulations.  We use the matching technique described in
\cite{rozo:methods} to match clusters to halos, and directly compute
the offset $R_s$ between the halo center and the BCG assigned to the
halo by the maxBCG cluster finding algorithm.  In the real Universe,
miscentering for our cluster population can occur for either of two
reasons --- the real BCG can be offset from the center of mass, or the
BCG can be misidenfied by the cluster finder.  In the mock catalogs,
there is always a bright galaxy at the center of the dark matter halo,
so we are neglecting here the first case.  Although this is not likely
to be precisely true in all cases, our results indicate that
miscentering due to misidentified BCGs dominates the effects we
discuss below.

For these catalogs, a richness-dependent fraction of
the BCGs appear to be accurately centered on their dark matter halos
($R_s \simeq 0$), while the rest are reasonably well described by a 2D
Gaussian distribution,
\begin{equation} 
P(R_s) = \frac{R_s}{\sigma_{s}^2} \exp(-\frac{1}{2} (R_s/\sigma_{s})^2) 
\label{eq:Rs} 
\end{equation} 
\noindent with $\sigma_s = 0.42~h^{-1}$ Mpc, independent of cluster
richness (see next section).  The resulting mean surface mass profile
for the miscentered clusters can be written
 
\begin{eqnarray} 
\Sigma_{NFW}^{s}(R) = \int dR_s P(R_s)~\Sigma_{NFW}(R | R_s)  
\end{eqnarray} 
 
\noindent and $\Delta\Sigma_{NFW}^{s}(R) = \overline{\Sigma^s}_{NFW}(<
R) - \overline{\Sigma^s}_{NFW}(R)$. We find that the mean shear
profile is not very sensitive to the shape of the distribution of
$R_s$, but it is sensitive to the effective scale length $\sigma_{s}$.
 
\begin{figure} 
\epsscale{1.1} 
\centering 
\plotone{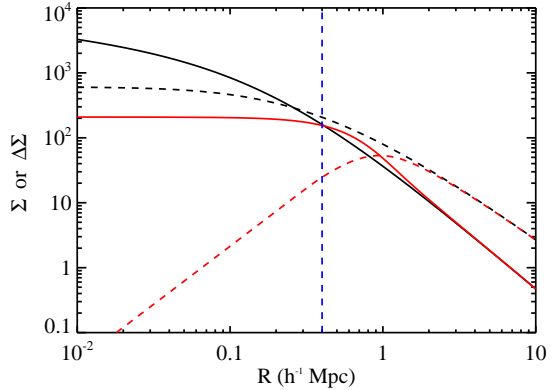} 
\caption{Effect of an offset between the BCG and the halo center on the projected mass  
profile $\Sigma(R)$ and the lensing signal $\Delta \Sigma(R)$.  
The black solid curve shows the $\Sigma(R)$  
profile for an NFW halo with $c_{200}=5$ and $r_{200}=1~ h^{-1}$ Mpc.  
The black dashed curve shows the corresponding  
$\Delta\Sigma(R)$ profile.  
The red curves show the resulting mean profiles  
when the distribution of randomly-oriented BCG-halo  
offsets is a 2D Gaussian with dispersion $\sigma_s = 0.42 h^{-1}$ Mpc  
(indicated by the blue vertical line).  
The red solid curve shows the smoothed $\Sigma^{s}(R)$ and the  
red dashed curve the smoothed $\Delta\Sigma^{s}(R)$ profile. Miscentering  
has the effect of making the $\Sigma^{s}(R)$ nearly flat, i.e., a mass sheet, at  
small scales. Although $\Sigma(R)$ and $\Sigma^{s}(R)$  
differ by only $10-30$\% near $r=\sigma_s$,  
$\Delta\Sigma$  and $\Delta\Sigma^{s}$ differ by an order of magnitude.  
For this example,  
$\Delta\Sigma^{s}(R)$ peaks at $r \simeq 2.5\sigma_s$; this  
behavior depends slightly on $c_{200}$. 
} 
\vspace*{0.3cm} 
\label{fig:convolve-NFW} 
\end{figure}

Figure \ref{fig:convolve-NFW} shows the effects of such miscentering
on the lensing signal for a cluster with an NFW profile. The effect on
$\Delta\Sigma(R)$ is much larger than on $\Sigma(R)$: the convolution
in Eqn. \ref{eq:conv} leads to a smoothing which essentially flattens
the $\Sigma^{s}(R)$ profile at small scales, creating a mass sheet
which causes little shear. While the $\Delta\Sigma_{NFW}(R)$ profile
is relatively flat at small scales, the smoothed
$\Delta\Sigma^{s}_{NFW}(R)$ profile is strongly suppressed at scales
$R \lesssim 2.5\sigma_s$.
 
In applying this model to the data in \S \ref{sec:results}, we include
$\ln(\sigma_s)$ as a model parameter, using its value from the mock
catalogs as the central value of a Gaussian prior probability
distribution. We assume that a fraction $p_c$ of the BCGs are
accurately centered on the dark matter halos, and that a fraction
$1-p_c$ follow the distribution of Eqn. \ref{eq:Rs}.  
The simulations are used to formulate a prior distribution
for $p_c$, as described in \S \ref{sec:halosummary}.
 
\begin{figure*} 
\epsscale{1.1} 
\plottwo{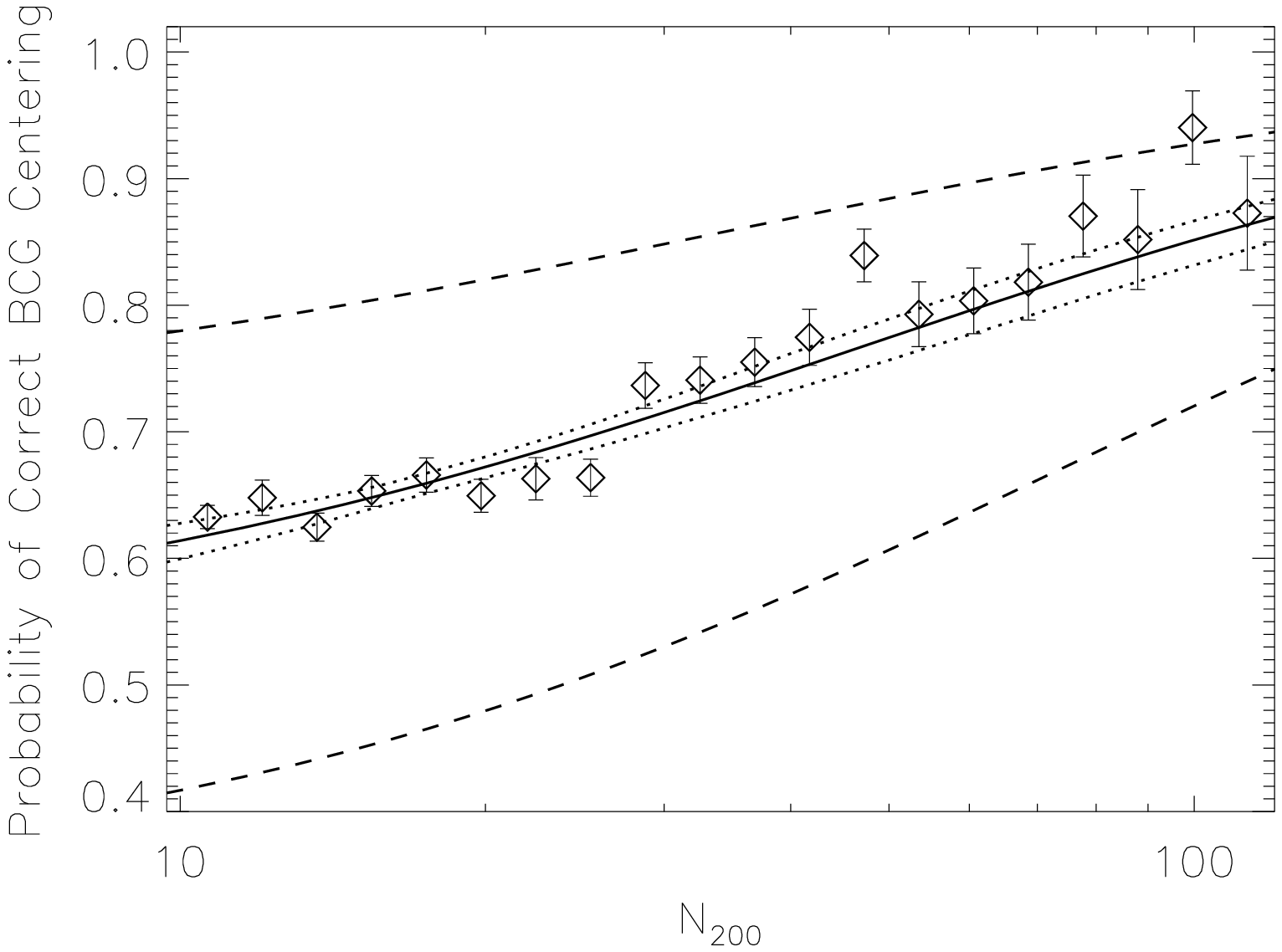}{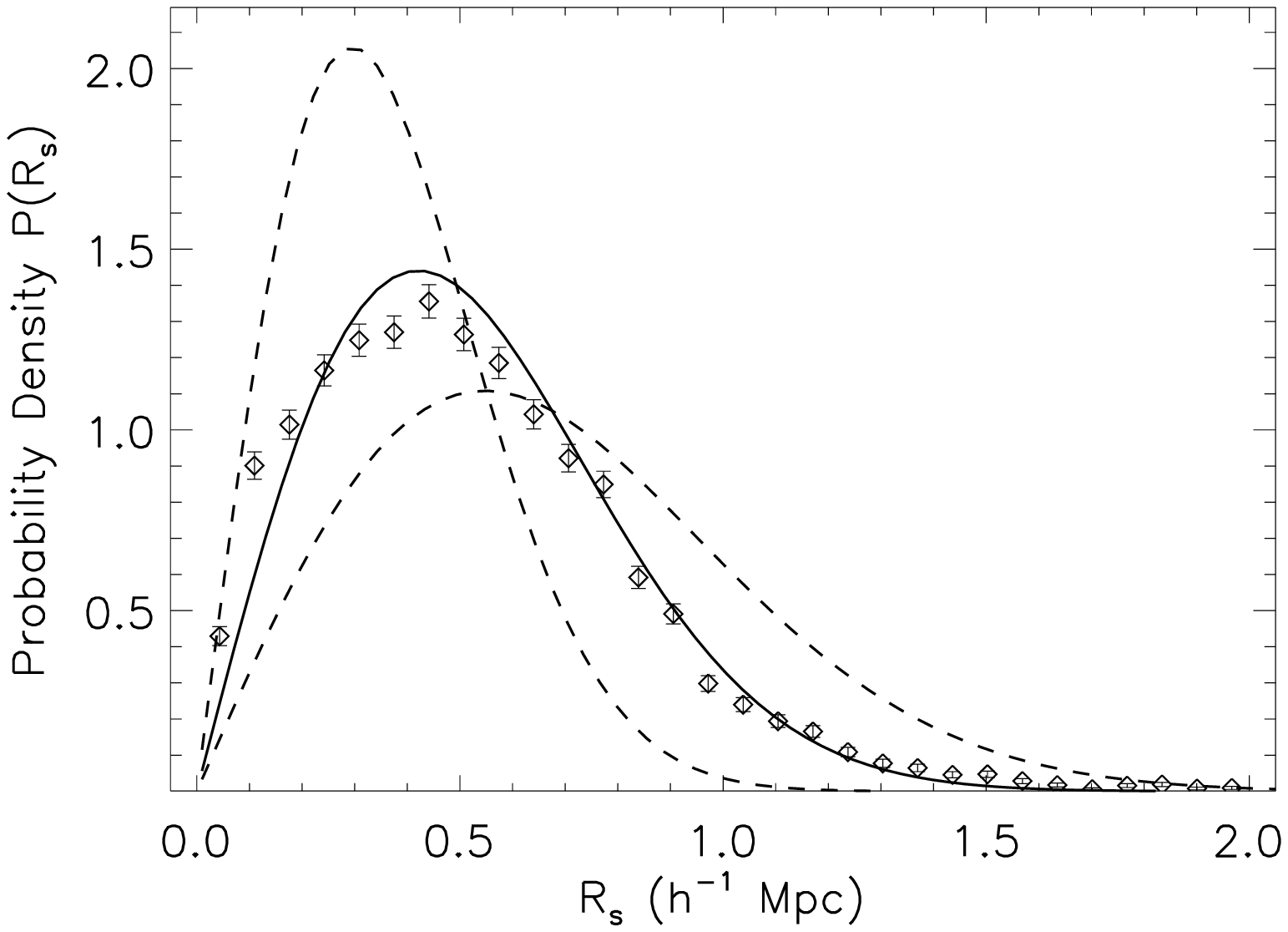} 
\caption{ {\bf Left:} The probability that a cluster is correctly
  centered as a function of cluster richness, $N_{200}$, in mock
  catalogs.  The diamonds with error bars are the measurements in the
  simulations, whereas the solid line is our best fit model (see text
  for details).  The dotted lines show the 95\% confidence ($2
  \sigma$) band from statistical uncertainties only.  The dashed line
  shows the more generous 95\% confidence region corresponding to the
  adopted 0.40 prior uncertainty on $q$, which are wider to allow for
  some possibility that there are differences in the probability of a
  cluster being correctly centered between our mocks and the real
  data.  {\bf Right:} The distribution of the projected radial offsets
  between a halos and clusters which are not correctly centered.
  Diamonds with error bars show the measurement in the mocks, while
  the solid line represents the best 2D Gaussian model, corresponding
  to a width $\sigma_S=0.42\ h^{-1}$ Mpc. The dashed lines are the two
  models that bound the much more generous 68\% confidence region of
  $\sigma_S$ with the adopted $30\%$ prior on $\sigma_S$.  }
\label{fig:miscenter-pars} 
\end{figure*}

We determine this fraction $p_c$ of correctly centered BCGs as a
function of $N_{200}$; this is shown in the left panel of Figure
\ref{fig:miscenter-pars}.  We can model this relation as $p_c(N_{200})
\equiv 1/(1+\exp(-q))$ with
\begin{equation} 
q = \ln (1.13 + 0.92~(N_{200}/20)).   
\label{eq:q-rich} 
\end{equation} 
The dotted lines show the statistical 95\% confidence bands recovered
in the simulations, whereas the dashed lines show the $95\%$ bands
corresponding to the much more generous $0.4$ prior on $q$ used in our
analysis as described in \S \ref{sec:halosummary}.  The right panel of
Figure \ref{fig:miscenter-pars} shows the miscentering distribution
$P(R_s)$.  The data from the simulations is roughly fit by a two
dimensional Gaussian of width $\sigma_R=0.42\ h^{-1}$ Mpc. Note that
because the mock catalogs place the BCG of a halo at the center of the
halo, the offset $R_s$ is identically zero if maxBCG assigns the
correct BCG to each cluster.
 
Our best fit model is shown as a solid line, while the dashed lines
show the models that bound the $68\%$ confidence regions corresponding
to the $30\%$ Gaussian prior on the parameter $\sigma_S$ used in \S
\ref{sec:halosummary} to fit the data.  It is clear that our adopted
priors are much more generous than the statistical noise in the
simulations. We choose this wider prior since there may be differences
between the mock catalogs and the real data.  The wider prior likely
can mostly account for real offsets between BCGs and the center of the
mass concentration.  Finally, we emphasize here that we are adopting
the same miscentering distribution for all richness bins. The
differences between the various richness bins in the mock data are
much smaller than the $30\%$ prior that we use.
 
\subsection{Neighboring mass concentrations}
\label{sec:twohalo}

The NFW profile is expected to be a good representation of the stacked
mass profiles on small to intermediate scales surrounding clusters, but on
large scales the lensing signal is dominated by neighboring mass concentrations, 
e.g., nearby halos and filaments. We model this contribution via the 
so-called two-halo term
\citep{seljak:halos,mandelbaum:diss-v-halo}, 

\begin{equation}
\rho_{2h}(r) = b(M_{200},z)~\Omega_m~\rho_{c, 0}~~(1+z)^3~\xi_l(r,z)
\end{equation}

\noindent where $\rho_{c, 0}$ is the critical density at the present epoch, and
$\xi_l(r,z)$ is the auto-correlation function of the mass in linear perturbation 
theory,
%the Fourier transform of the power spectrum, 
evaluated at the redshift of the clusters.  Here, $b(M_{200},z)$ is
the linear bias parameter for dark matter halos, which has a predicted
dependence upon halo mass and redshift
\citep{sheth-tormen:cluster-bias,seljak-warren:halo-bias}.
 
The shape of the linear correlation function is determined by the
cosmological parameters $n_s$, $h$, and $\Omega_m$ for a flat LCDM
model and is constrained by observations of galaxy clustering
\citep{eisenstein:wiggles,zehavi:departure}. The linear correlation
function can be expressed as

\begin{equation}
\xi_l(r,z) = D(z)^2~\sigma_8^2~\xi_l((1+z)~r)~,
\end{equation}

\noindent where $\xi_l(r)$ with a single argument 
is the linear correlation function evaluated at $z=0$
and normalized to $\sigma_8=1$. The presence of the factor of
$(1+z)$ in the above expression converts the \emph{physical} distance $r$
into \emph{comoving} units. All distances in this paper are in \emph{physical} 
not \emph{comoving} units. The linear growth factor satisfies

\begin{equation}
D(a) \propto H(a) ~\int da^{\prime} \left[ H(a^{\prime}) a^{\prime} \right]^{~-3}~,
\label{eq:linear}
\end{equation}

\noindent with $a=1/(1+z)$; $D$ is normalized to unity at $a=1$ ($z=0$). 
We can therefore express the two-halo contribution to the density as

\begin{equation}
\rho_{2h}(r) = B\rho_{c,0}~(1+z)^3~\xi_l((1+z)~r)~,
\label{eq:r2h2}
\end{equation}

\noindent where we have defined an effective bias parameter 

\begin{equation}
B \equiv b(M_{200},z)~\Omega_m~\sigma_8^2~D(z)^2~.
\label{eq:biasdef}
\end{equation}

The contribution of the two-halo term to the lensing signal, for 
fixed values of $n_s$, $h$, and $\Omega_m$, can be 
written as $\Delta \Sigma(R;B)=B \Delta \Sigma_l$, where, as before, 
$\Delta \Sigma_l(R) = \overline{\Sigma}_l(< R) - \overline{\Sigma}_l(R)$, 
and 

\begin{eqnarray}
\Sigma_l(R) &=& (1+z)^3 \rho_{c,0} \int dy \xi_l \left((1+z)\sqrt{y^2+R^2}\right) \nonumber \\
&=& (1+z)^2 \rho_{c,0} ~ W \left( (1+z)R \right)
\end{eqnarray}

\noindent with 

\begin{equation}
W(R) \equiv \int dy~ \xi_l(\sqrt{y^2+R^2})~.
\end{equation}

\subsection{Summary of halo model and parameter priors for $\Delta\Sigma$ fits}
\label{sec:halosummary}
Combining the results from sections \ref{sec:bcg} through \ref{sec:twohalo}, 
we can write down the model for the lensing signal $\Delta\Sigma$ thus far,

\begin{eqnarray}
& & \Delta\Sigma(R) =  \nonumber \\ 
& &  {M_0\over \pi R^2} + p_c\Delta\Sigma_{NFW}(R) + (1-p_c)\Delta\Sigma_{NFW}^{s}(R) 
+ B\Delta\Sigma_l \nonumber \\
\label{eq:full-model-1}
\end{eqnarray}

\noindent where, sequentially, the terms come from the BCGs, 
the halos centered on the BCGs, the halos not centered on the BCGs, and the 
neighboring halos. 

There are two further effects to consider. This model assumes a
constant halo mass where, in reality, the signal will be averaged over
the distribution of halos masses for each richness bin.  The other
effect that we will consider is the non-linear shear effect that is
discussed in \cite{mandelbaum:groups}. We will treat this non-linear
contribution first and then integrate the full signal over the
distribution of masses.

The average tangential ellipticities do not trace the shear exactly
but rather trace the \emph{reduced shear}, $g \equiv
\gamma/(1-\kappa)$.  Let $e_{ij}$ be the $i$-th source galaxy around
cluster $j$ for some radial bin.  As shown in \cite{mandelbaum:groups}
an estimator for $\Delta\Sigma$ formed from a weighted average of
ellipticities and identical halos has a second order contribution

\begin{eqnarray}
\widehat{\Delta\Sigma} & = &  \sum_{ij} W_{ij}~e_{ij} \nonumber \\
& = & \Delta\Sigma + \Delta\Sigma~\Sigma~\mathcal{L}_Z 
\end{eqnarray}
with
\begin{equation}
\mathcal{L}_Z = \left<\Sigma_{crit}^{-3}\right>/\left<\Sigma_{crit}^{-2}\right>
\end{equation}

This differs from the $\left<\Sigma_{crit}^{-2}\right>/\left<\Sigma_{crit}^{-1}\right>$ in \cite{mandelbaum:groups} in that our weighting has an 
explicit factor of $\Sigma_{crit}^{-1}$.

\begin{equation}
W_{ij} = \frac{1}{2R} \frac{\sigma_{ij}^{-2} \Sigma_{crit}^{-1}(i,j)}{\sum_{kl} \sigma_{kl}^{-2} \Sigma_{crit}^{-2}(k,l)}
\end{equation}
where $R$ is the shear responsivity and $\sigma_{kl}^{2}$ are the estimates of variances on source ellipticities.

Using the photometric redshifts for the source galaxies and the maxBCG
photometric estimates for the cluster redshifts, we find
$\mathcal{L}_Z = 1.40 \times 10^{-4} ~h^{-1}~{\rm pc}^2/M_{\sun}$.
This quantity varies only a few percent across different cluster
samples and different radial bins; a variation we ignore.

For the last step we need to consider that within any richness bin, there will be scatter in mass. So we need to integrate this
expression over the probability distribution of halos masses $P(M_{200})$.
Here, $\left<~\right>$ indicates averaging over $P(M_{200})$.

\begin{equation}
\widehat{\left<\Delta\Sigma\right>} =  \left<\Delta\Sigma\right> + \left<\Delta\Sigma~\Sigma\right>~\mathcal{L}_Z
\end{equation}

We will use a log-normal distribution of $M_{200}$ at fixed richness with a variance in $\ln M_{200}$ given by
$V_M$ which is our last model parameter.

For the first term $\left< \Delta\Sigma \right>$ we can integrate Eqn. \ref{eq:full-model-1} over $P(M_{200})$. Corresponding to 
Eqn. \ref{eq:full-model-1}, there is a three-term expression for $\Sigma$ (the point-mass doesn't contribute).
So our second order correction has 12 terms that need to be integrated over $P(M_{200})$. Most of these pairs $(i,j)$ do not
contribute since $\left<\Delta\Sigma_i~\Sigma_j\right>(R)~\mathcal{L}_Z \ll \left<\Delta\Sigma\right>(R)$ at all scales.

Only two of these terms make meaningful contributions at the smallest scales,
\begin{eqnarray} 
\Delta\Sigma_{NL} = \mathcal{L}_Z ~ \times && \nonumber \\
\left[  p_c^2 \left<\Delta\Sigma_{NFW}~\Sigma_{NFW}\right> + \frac{p_c M_0}{\pi R^2} \left<\Sigma_{NFW}\right> \right]. &&
\end{eqnarray} 

With this last expression we can write down our full model for our data 
where, again, $\left<~\right>$ indicates averaging over $P(M_{200})$.

\begin{eqnarray}
& & \widehat{<\Delta\Sigma(R)>} =  \nonumber \\
& &  {M_0\over \pi R^2} + p_c\left<\Delta\Sigma_{NFW}\right>(R) \nonumber \\
& & (1-p_c)\left<\Delta\Sigma_{NFW}^{s}\right>(R) + B\Delta\Sigma_l + \Delta\Sigma_{NL}\nonumber \\
\label{eq:full-model}
\end{eqnarray}

This model has seven parameters: the BCG point mass $M_0$; the two NFW halo 
parameters $r_{200}$ and $c_{200}$; the scatter in the mass--richness relation $V_M$;
the halo miscentering width $\sigma_s$and the halo 
centering fraction $p_c$; the linear bias amplitude $B = b(M_{200}) \Omega_m \sigma_8^2 D^2(z)$
which should also be thought of as the \emph{average} over the mass distribution.

Since we will integrate over $P(M_{200})$ we need to be specific about what we mean by our parameter $r_{200}$.
We take this to be the $r_{200}$ corresponding to the \emph{average} $M_{200}$. For a log-normal distribution
$<M_{200}> = \exp( V_M/2 + \mu)$ where $\mu$ is the average $\ln(M_{200})$.

\cite{becker:vel} measures the variance of the logarithm of the galaxy velocity dispersion as
$Var(\sigma_v) = 0.0963 - 0.0241 (N_{200}/25)$ and \cite{evrard:mass-vel} determines the
scaling $M_{200} \propto \sigma_v^{\lambda}$ with $\lambda=2.98$. This results in 

\begin{equation}
V_M = 0.855 - 0.214 \ln (N_{200}/25).
\label{eq:VM-rich}
\end{equation}
We allow for an uncertainty of 0.60 in $V_M$ in our prior (i.e. a 30\% 
uncertainty for the scatter). This log-normal model also seems consistent with our mock catalogs.

\begin{deluxetable*}{cccccc}
\tablecaption{Halo Model Parameters for $\Delta \Sigma$ fits}
\tablewidth{0pt}
\tablehead{
\colhead{Param \#} & \colhead{Parameter} & \colhead{Description} & \colhead{Prior-mean} & \colhead{Prior-sigma} & \colhead{Note}
}
\startdata
1 & $\ln(r_{200})$ & $r_{200}$ radius  & -0.693 & 1.5 & weak prior \\
2 & $\ln(c_{200})$ & concentration & 1.386 & 3 & weak prior \\
3 & $B$ & bias amplitude & 0.5 & 4.0 & weak prior \\
4 & $q$ & miscentering parameter & see text & 0.4 & strong prior \\
5 & $\ln(\sigma_S)$ & miscentering width & -0.868 & 0.3 & strong prior \\
6 & $M_0$ & point mass $(10^{12}~h^{-1} M_{\sun})$& 0 & 2.5 & weak prior \\
7 & $V_M$ & variance of $\ln(M_{200})$ & see text & 0.6 & strong prior \\
\enddata
\tablecomments{Parameters in the model
  (Eqn. \ref{eq:full-model}) for $\Delta\Sigma$. The mean and
  standard deviation for the Gaussian prior distribution are given as well as a
  brief description.  
}
\label{tab:parameters}
\end{deluxetable*}

Table \ref{tab:parameters} lists all seven model parameters used in the
fits, including the information about the prior distributions. To
enforce positivity, logarithms are used for all parameters except $B$
and $M_0$. Each prior distribution is taken to be a Gaussian with mean
and standard deviation as indicated in the table. In addition $M_0$ is 
forced to be positive since it is not at all constrained on the lower end.
Here, a ``weak prior'' means that neither the best-fit parameters nor the estimated
parameter errors change significantly if the standard deviation of the
prior distribution is increased.  Since the parameter $p_c$ is
constrained to lie in the range $[0,1]$, we use the transformed
parameter $q \equiv \ln[p_c/(1-p_c)]$ which has range $[-\infty, +\infty]$ and
can thus be assigned a Gaussian prior.  The prior mean of
$q$ and $V_M$ vary with richness as described in Eqn. \ref{eq:q-rich} and \ref{eq:VM-rich}.

To fit the measured $\Delta\Sigma$ profiles with the model, 
we use a Markov Chain Monte Carlo (MCMC). MCMC is useful for efficiently 
calculating likelihoods in multi-dimensional parameter spaces. MCMC methods generate 
``chains'' or sequences in the parameter space that 
represent a fair sampling from the full posterior probability distribution. 
Thus, they allow one to visualize the likelihood surface and
see degeneracies between parameters without assuming that the 
errors are normally distributed (as in the Fisher matrix method). 
It is also straightforward to include priors on parameters in the MCMC 
approach. Our MCMC routine uses the Metropolis-Hastings algorithm with 
Gaussian transition functions. The total number of steps is 100,000, and 
we discard a burn-in period of the first 1000 steps. Runs of varying length show
that convergence of the posterior distribution is reached before 10,000 steps
and longer runs such as our 100,000 step run improve the sampling but do not
affect the sample mean or variances by meaningful amounts.

\subsection{Systematic errors}
\label{sec:sys}

There are two major sources of systematic error in any weak lensing measurements: shear calibration error and 
errors associated with photo-$z$ biases. 
The shear calibration error of the shear estimation methods that we use for the SDSS were tested as part of the 
Shear TEsting Program II (STEP2 ; \cite{step2}) and found to be less
than a percent (for the RM method in STEP2). However, we
allow for a 3\% error in shear calibration since the STEP simulation error
may not represent the full calibration error when these methods are
applied to real data --- e.g., the PSF modeling of SDSS was not tested in STEP.

The dominant systematic error is that associated with biases in the photometric redshift
distribution. We use a neural network based method (Cunha et al. In preparation) which uses 
a training set of spectroscopic redshifts from the SDSS, CNOC2 \citep{yee:cnoc2} and
CFRS \citep{lilly:cfrs}. See Paper I for details.

Although it is difficult to estimate the residual photo-$z$ bias, 
we will assume that the amplitude of the resultant masses is 
uncertain at the level of 7\% and we will include this in our errors of
the zero-point of the mass--richness relation. Further improvements in photo-$z$ 
calibration should be able to reduce this overall error by as much as a few percent.

Another source of systematic errors is model dependency. 
The priors that we have chosen are considered to be independent between richness 
bins and so combining 12 to 16 bins of data reduces the effective width of these priors by about 4 when considering 
averaged quantities such as the mass--richness relation (see \S \ref{sec:massrich}).
However if the prior means for these quantities such as $q$ and $V_M$ are shifted \emph{systematically} from their true values,
the effect of these maginalizations may not fully account for this. By experimenting with different values for these
prior means we can estimate the possible level of additional systematic errors. For the mass--richness relation we estimate
that this will contribute an additional systematic error of 10\%. The concentration, $c_{200}$, is more affected by shifts
in these nuisance parameters, particularly $q$ and $M_0$. We allow for a 30\% systematic error on the amplitude of the
$c_{200}$--$M_{200}$ relation \S \ref{sec:conc}. The bias parameter, $B$ is less affected by these nuisance parameters but as
we shall see in \S \ref{sec:bias}, some knowledge of $V_M$ is required to compare it with theoretical predictions.

\section{Results of halo model fits}
\label{sec:results}

\begin{figure*}
\epsscale{1.0}
\plotone{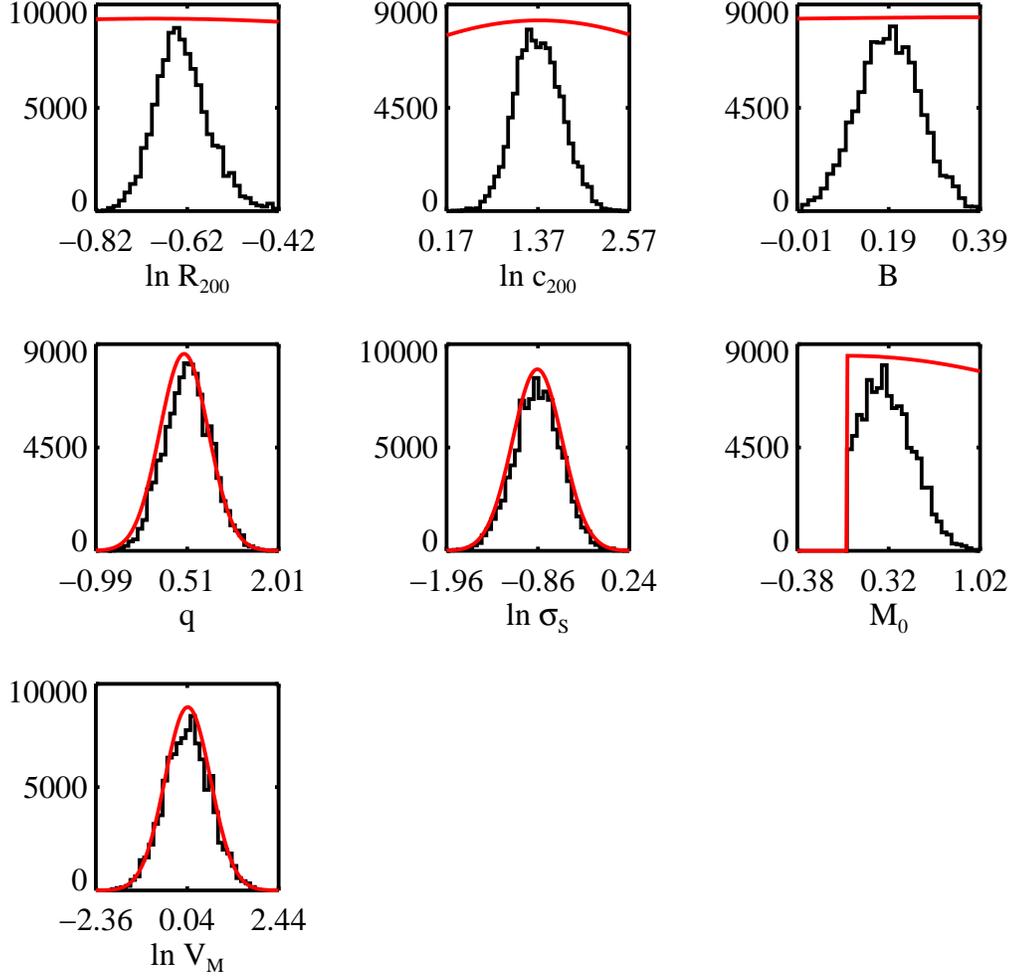}
\caption{This shows the one-dimensional marginal posteriors for the 7 parameters
for the seventh $N_{200}$ richness bin. Most resemble a normal distribution
The red lines indicate the prior normal distribution for each (with arbitrary normalization).
The first three $\ln R_{200}$, $\ln c_{200}$ and $B$ have uninformative priors whereas $q$, $\ln \sigma_S$ and $\ln V_M$.
have constraining priors; i.e. the posterior resembles the prior so the data are uninformative for these.
The prior for $M_0$ is constrained to be positive but is largely uninformative beyond that.
\label{fig:param_hist}}
\end{figure*}

Figure \ref{fig:param_hist} shows the result of an MCMC run for the
seventh $M_{200}$ richness bin. These are the one-dimensional marginal
posteriors for the 7 parameters.  Most resemble a normal distribution
with the exception of $M_0$ which is constrained to be strictly
positive.  The red lines indicate the prior normal distribution for
each.  The first three $\ln R_{200}$, $\ln c_{200}$ and $B$ have
uninformative priors whereas $q$, $\ln \sigma_S$ and $\ln V_M$ have
constraining priors. That is, in the later case, the posterior
resembles the prior; the data are uninformative for these three.

\begin{figure*}
\epsscale{1.0}
\plotone{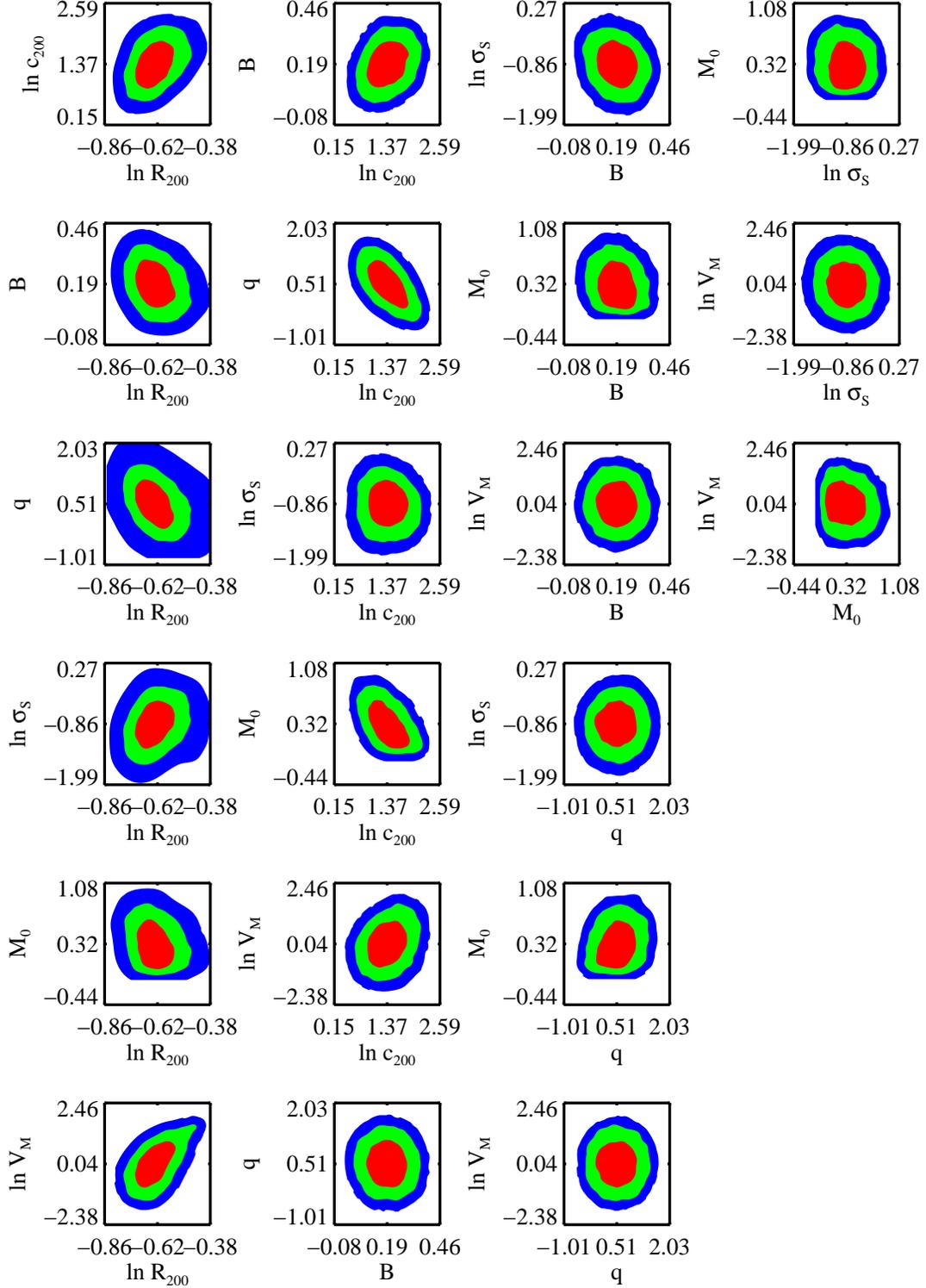}
\caption{The results of the MCMC chain for the seventh $N_{200}$ richness bin.
This shows the marginal posterior distributions for all 21 pairs of parameters.
The red region is the 68\% (1 $\sigma$) confidence region; green is
the 95\% (2 $\sigma$) confidence region and blue is the 99\% (3 $\sigma$) confidence region.  
Although none of these parameters appear to be strongly degenerate at 
these noise levels, there is some correlation. $R_{200}$ is correlated with 
both $q$ and $V_M$ and $c_{200}$ is correlated with $q$ and $M_0$. 
These contours also allow for an estimate of how the best fit parameters might 
be biased if we have systematically misestimated our nuisance parameter priors. 
\label{fig:covar}}
\end{figure*}

Figure \ref{fig:covar} shows the marginal posterior distributions for all 21 pairs of parameters for the same bin.
The red region is the 68\% (1 $\sigma$) confidence region; green is
the 95\% (2 $\sigma$) confidence region and blue is the 99\% (3 $\sigma$) confidence region.
Although none of these parameters appear to be strongly degenerate at
these noise levels, there is some correlation. $R_{200}$ is correlated with
both $q$ and $V_M$ and $c_{200}$ is correlated with $q$ and $M_0$. These contours also allow
for an estimate of how the best fit parameters might be biased if we have systematically 
misestimated our nuisance parameter priors.  If the shot noise were significantly
smaller, these correlations with nuisance parameters would become more dominant sources of error, and so modeling
the effect of these (and possibly other) parameters will become a more critical issue for future experiments.

%\begin{figure*}
%\epsscale{1.1}
%\plottwo{Plots/Ngal200_DS_hfits.ps}{Plots/iLum200_DS_hfits.ps}
%\caption{Model fits to $\Delta\Sigma(R)$ for the 12 $N_{200}$ richness bins ({\it left panel}) and 
%for the 16 $L_{200}$ luminosity bins ({\it right panel}). The model components are the 
%NFW halo profile (green), miscentered halo component (orange), the central BCG (red), 
%neighboring halos (blue); the non-linear contribution (purple dashed). 
%The magenta curves show the sum of these components for the best-fit models in each bin.
%}
%\label{fig:ds-fit}
%\end{figure*}

\begin{figure*}
\epsscale{1.1}
\plotone{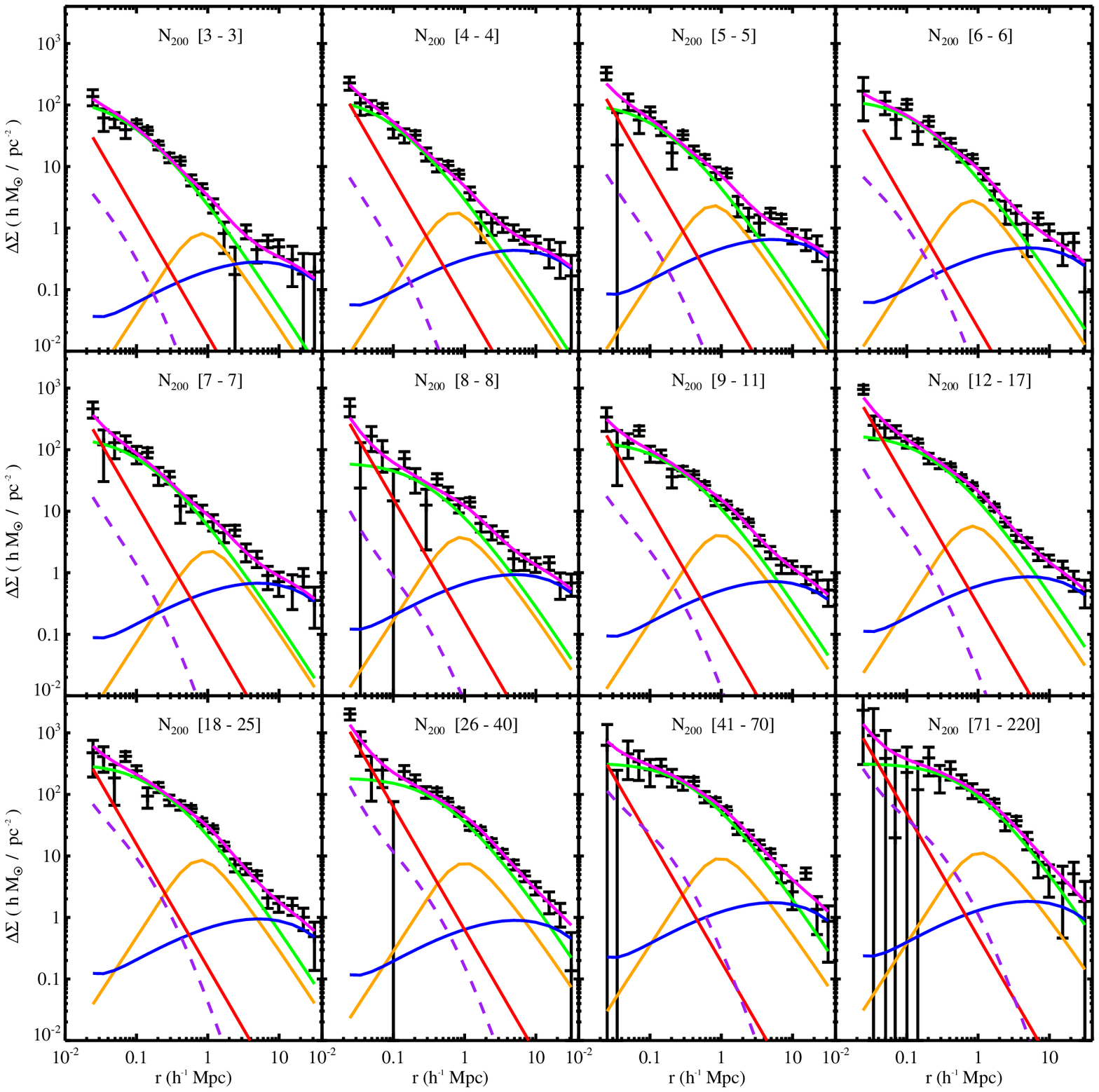}
\caption{Model fits to $\Delta\Sigma(R)$ for the 12 $N_{200}$ richness bins 
The model components are the
NFW halo profile (green), miscentered halo component (orange), the central BCG (red),
neighboring halos (blue); the non-linear contribution (purple dashed).
The magenta curves show the sum of these components for the best-fit models in each bin.
}
\label{fig:ds-fit-n200}
\end{figure*}

\begin{figure*}
\epsscale{1.1}
\plotone{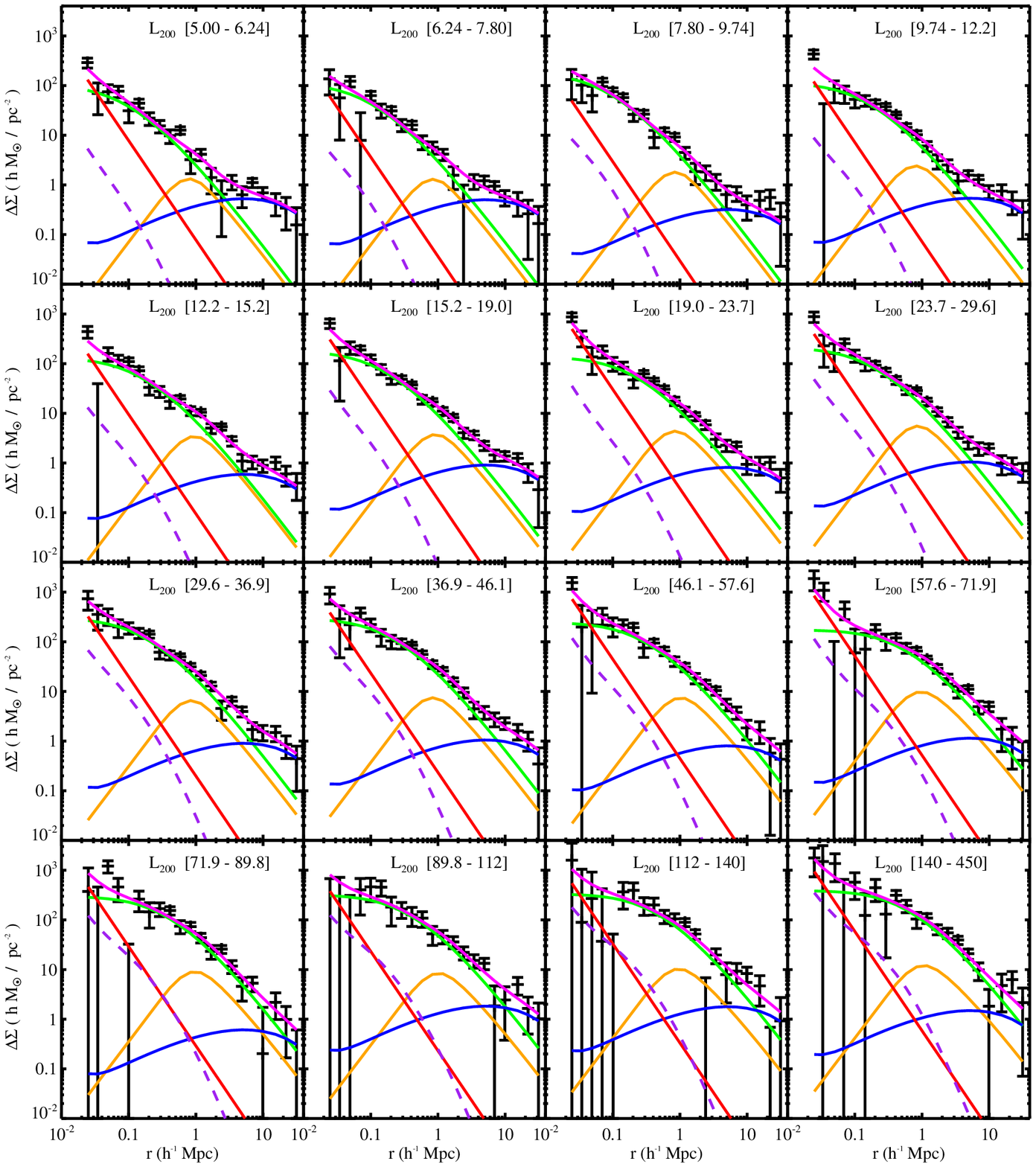}
\caption{Model fits to $\Delta\Sigma(R)$ for the 16 $L_{200}$ luminosity bins. The model components are the
NFW halo profile (green), miscentered halo component (orange), the central BCG (red),
neighboring halos (blue); the non-linear contribution (purple dashed).
The magenta curves show the sum of these components for the best-fit models in each bin.
}
\label{fig:ds-fit-l200}
\end{figure*}

The results of fitting this model to the $\Delta \Sigma$ profiles in
the 12 $N_{200}$ richness and 16 $L_{200}$ luminosity bins are shown
in Figures \ref{fig:ds-fit-n200} and \ref{fig:ds-fit-l200}. In each panel, the green curve shows the
NFW halo profile, the blue curve indicates the two-halo term, the red curve is the BCG point mass
term, the orange curve is the smoothed (miscentered) NFW halo
component, and the purple dashed curve shows the non-linear correction.
The magenta curve shows the sum of these terms.  One
can see that the model does a good job of fitting all of the features
in the shear profiles, the most prominent of which is the one-halo to
two-halo transition, which usually occurs near $r_{200}$. The best fit
parameters for $r_{200}$, $c_{200}$ and $B$, properly marginalized
over the nuisance parameters, are shown in Tables
\ref{tab:mass-rich-ngal} and \ref{tab:mass-rich-lum}. We show the
values for mass and concentration converted to other mass definitions
in Tables \ref{tab:mass-conv-ngal} and \ref{tab:mass-conv-lum}.
The method of conversion is discussed in the Appendix.

\begin{deluxetable*}{ccccc}
\tablecaption{Best Fit Parameters: $N_{200}$ Bins}
\tablewidth{0pt}
\tablehead{
\colhead{$<N_{200}>$} &
\colhead{$M_{200}$ ( $10^{12}h^{-1}M_{\sun}$ )} &
\colhead{$r_{200}$ ($h^{-1}$ Mpc)} &
\colhead{$c_{200}$} &
\colhead{B}
}
\startdata
      3.00 &       6.37 $\pm$       1.04 &       0.28 $\pm$      0.015 &       5.78 $\pm$       1.35 &       0.07 $\pm$       0.03 \\
      4.00 &       9.77 $\pm$       1.80 &       0.32 $\pm$      0.020 &       6.17 $\pm$       2.29 &       0.11 $\pm$       0.04 \\
      5.00 &      14.63 $\pm$       2.90 &       0.37 $\pm$      0.024 &       4.45 $\pm$       1.58 &       0.17 $\pm$       0.05 \\
      6.00 &      21.35 $\pm$       3.66 &       0.42 $\pm$      0.024 &       4.33 $\pm$       1.12 &       0.13 $\pm$       0.06 \\
      7.00 &      23.31 $\pm$       5.56 &       0.43 $\pm$      0.034 &       5.77 $\pm$       2.35 &       0.18 $\pm$       0.07 \\
      8.00 &      27.86 $\pm$       6.97 &       0.46 $\pm$      0.038 &       2.34 $\pm$       1.01 &       0.25 $\pm$       0.09 \\
      9.82 &      44.14 $\pm$       7.96 &       0.53 $\pm$      0.032 &       3.97 $\pm$       1.21 &       0.19 $\pm$       0.07 \\
     13.91 &      60.01 $\pm$       8.45 &       0.59 $\pm$      0.028 &       4.22 $\pm$       1.12 &       0.23 $\pm$       0.08 \\
     20.78 &      95.96 $\pm$      12.58 &       0.69 $\pm$      0.030 &       5.82 $\pm$       1.49 &       0.25 $\pm$       0.10 \\
     31.09 &     167.76 $\pm$      23.39 &       0.83 $\pm$      0.039 &       2.95 $\pm$       0.66 &       0.24 $\pm$       0.13 \\
     50.27 &     252.06 $\pm$      35.28 &       0.95 $\pm$      0.044 &       4.01 $\pm$       0.86 &       0.46 $\pm$       0.20 \\
     92.18 &     568.81 $\pm$      87.75 &       1.25 $\pm$      0.064 &       2.92 $\pm$       0.76 &       0.48 $\pm$       0.36 \\
\enddata
\tablecomments{
This shows the best fit parameters of interest from the MCMC for the $N_{200}$ richness bins.We have marginalized over the four nuisance parameters.
}
\label{tab:mass-rich-ngal}
\end{deluxetable*}

\begin{deluxetable*}{ccccc}
\tablecaption{Best Fit Parameters: $L_{200}$ Bins}
\tablewidth{0pt}
\tablehead{
\colhead{$<L_{200}>$} &
\colhead{$M_{200}$ ( $10^{12}h^{-1}M_{\sun}$ )} &
\colhead{$r_{200}$ ($h^{-1}$ Mpc)} &
\colhead{$c_{200}$} &
\colhead{B}
}
\startdata
      5.59 &       7.87 $\pm$       1.84 &       0.30 $\pm$      0.023 &       5.31 $\pm$       2.39 &       0.14 $\pm$       0.04 \\
      6.97 &       9.19 $\pm$       1.91 &       0.32 $\pm$      0.022 &       5.21 $\pm$       1.79 &       0.13 $\pm$       0.04 \\
      8.69 &      13.62 $\pm$       2.45 &       0.36 $\pm$      0.022 &       6.86 $\pm$       1.88 &       0.08 $\pm$       0.04 \\
     10.84 &      18.23 $\pm$       3.22 &       0.40 $\pm$      0.023 &       4.20 $\pm$       1.20 &       0.14 $\pm$       0.05 \\
     13.53 &      29.65 $\pm$       5.99 &       0.47 $\pm$      0.031 &       4.77 $\pm$       1.73 &       0.16 $\pm$       0.06 \\
     16.89 &      37.44 $\pm$       6.36 &       0.50 $\pm$      0.029 &       5.20 $\pm$       1.70 &       0.24 $\pm$       0.07 \\
     21.06 &      41.79 $\pm$       7.27 &       0.52 $\pm$      0.030 &       3.88 $\pm$       1.19 &       0.22 $\pm$       0.07 \\
     26.31 &      59.58 $\pm$       9.34 &       0.59 $\pm$      0.031 &       4.99 $\pm$       1.47 &       0.28 $\pm$       0.09 \\
     32.89 &      78.32 $\pm$      11.88 &       0.64 $\pm$      0.033 &       6.01 $\pm$       1.62 &       0.24 $\pm$       0.10 \\
     40.95 &      97.25 $\pm$      14.51 &       0.69 $\pm$      0.034 &       5.41 $\pm$       1.47 &       0.28 $\pm$       0.11 \\
     51.19 &     141.43 $\pm$      23.27 &       0.79 $\pm$      0.043 &       4.16 $\pm$       1.22 &       0.21 $\pm$       0.12 \\
     64.08 &     204.05 $\pm$      33.23 &       0.89 $\pm$      0.048 &       2.67 $\pm$       0.75 &       0.30 $\pm$       0.18 \\
     79.89 &     210.75 $\pm$      35.03 &       0.90 $\pm$      0.050 &       4.09 $\pm$       1.13 &       0.16 $\pm$       0.12 \\
     98.69 &     235.24 $\pm$      47.69 &       0.93 $\pm$      0.063 &       4.11 $\pm$       1.41 &       0.48 $\pm$       0.27 \\
    124.59 &     327.90 $\pm$      62.23 &       1.04 $\pm$      0.066 &       3.75 $\pm$       1.13 &       0.47 $\pm$       0.31 \\
    184.65 &     610.42 $\pm$      99.89 &       1.28 $\pm$      0.070 &       3.45 $\pm$       0.90 &       0.39 $\pm$       0.31 \\
\enddata
\tablecomments{
This shows the best fit parameters of interest from the MCMC for the $L_{200}$ richness bins.We have marginalized over the four nuisance parameters.
}
\label{tab:mass-rich-lum}
\end{deluxetable*}

\begin{deluxetable*}{ccccccccc}
\tablecaption{Mass Richness: $N_{200}$ Bins}
\tablewidth{0pt}
\tablehead{
\colhead{$<N_{200}>$} &
\colhead{$M_{200}$} &
\colhead{$c_{200}$} &
\colhead{$M_{180b}$} &
\colhead{$c_{180b}$} &
\colhead{$M_{vir}$} &
\colhead{$c_{vir}$} &
\colhead{$M_{500}$} &
\colhead{$c_{500}$}
}
\startdata
      3.00 &       6.37 &       5.78 &       8.27 &       8.72 &       7.41 &       7.26 &       4.72 &       3.85 \\
      4.00 &       9.77 &       6.17 &      12.58 &       9.28 &      11.30 &       7.74 &       7.31 &       4.12 \\
      5.00 &      14.63 &       4.45 &      19.66 &       6.80 &      17.35 &       5.64 &      10.34 &       2.92 \\
      6.00 &      21.35 &       4.33 &      28.82 &       6.63 &      25.39 &       5.49 &      15.01 &       2.84 \\
      7.00 &      23.31 &       5.77 &      30.25 &       8.71 &      27.08 &       7.25 &      17.24 &       3.84 \\
      8.00 &      27.86 &       2.34 &      42.04 &       3.71 &      35.40 &       3.03 &      16.89 &       1.46 \\
      9.82 &      44.14 &       3.97 &      60.39 &       6.09 &      52.91 &       5.04 &      30.49 &       2.58 \\
     13.91 &      60.01 &       4.22 &      81.34 &       6.45 &      71.54 &       5.35 &      41.97 &       2.76 \\
     20.78 &      95.96 &       5.82 &     124.42 &       8.78 &     111.44 &       7.31 &      71.09 &       3.88 \\
     31.09 &     167.76 &       2.95 &     241.65 &       4.60 &     207.35 &       3.78 &     108.26 &       1.88 \\
     50.27 &     252.06 &       4.01 &     344.24 &       6.16 &     301.85 &       5.10 &     174.52 &       2.62 \\
     92.18 &     568.81 &       2.92 &     820.81 &       4.56 &     703.78 &       3.75 &     366.17 &       1.86 \\
\enddata
\tablecomments{
Maximum likelihood mean halo mass and concentration parameters for each richness bin converted from our $200 \rho_{c}$ definition of virial mass  into three other common definitions. The unit of mass is $10^{12}h^{-1}M_{\sun}$.
}
\label{tab:mass-conv-ngal}
\end{deluxetable*}

\begin{deluxetable*}{ccccccccc}
\tablecaption{Mass Richness: $L_{200}$ Bins}
\tablewidth{0pt}
\tablehead{
\colhead{$<L_{200}>$} &
\colhead{$M_{200}$} &
\colhead{$c_{200}$} &
\colhead{$M_{180b}$} &
\colhead{$c_{180b}$} &
\colhead{$M_{vir}$} &
\colhead{$c_{vir}$} &
\colhead{$M_{500}$} &
\colhead{$c_{500}$}
}
\startdata
      5.59 &       7.87 &       5.31 &      10.32 &       8.04 &       9.20 &       6.69 &       5.74 &       3.52 \\
      6.97 &       9.19 &       5.21 &      12.09 &       7.89 &      10.76 &       6.56 &       6.68 &       3.45 \\
      8.69 &      13.62 &       6.86 &      17.32 &      10.29 &      15.63 &       8.59 &      10.34 &       4.61 \\
     10.84 &      18.23 &       4.20 &      24.72 &       6.43 &      21.74 &       5.32 &      12.74 &       2.74 \\
     13.53 &      29.65 &       4.77 &      39.47 &       7.26 &      34.98 &       6.03 &      21.23 &       3.14 \\
     16.89 &      37.44 &       5.20 &      49.25 &       7.88 &      43.85 &       6.55 &      27.22 &       3.44 \\
     21.06 &      41.79 &       3.88 &      57.37 &       5.97 &      50.20 &       4.93 &      28.73 &       2.52 \\
     26.31 &      59.58 &       4.99 &      78.81 &       7.58 &      70.01 &       6.30 &      43.00 &       3.30 \\
     32.89 &      78.32 &       6.01 &     101.14 &       9.06 &      90.74 &       7.55 &      58.32 &       4.01 \\
     40.95 &      97.25 &       5.41 &     127.27 &       8.18 &     113.56 &       6.81 &      71.18 &       3.59 \\
     51.19 &     141.43 &       4.16 &     192.06 &       6.38 &     168.80 &       5.28 &      98.64 &       2.72 \\
     64.08 &     204.05 &       2.67 &     299.66 &       4.19 &     255.11 &       3.44 &     128.33 &       1.68 \\
     79.89 &     210.75 &       4.09 &     286.93 &       6.28 &     251.92 &       5.19 &     146.50 &       2.67 \\
     98.69 &     235.24 &       4.11 &     320.05 &       6.30 &     281.08 &       5.22 &     163.68 &       2.68 \\
    124.59 &     327.90 &       3.75 &     452.69 &       5.78 &     395.19 &       4.77 &     223.80 &       2.43 \\
    184.65 &     610.42 &       3.45 &     854.38 &       5.35 &     741.71 &       4.41 &     409.17 &       2.23 \\
\enddata
\tablecomments{
Maximum likelihood mean halo mass and concentration parameters for each luminosity bin converted from our $200 \rho_{c}$ definition of virial mass into three other common definitions. The unit of mass is $10^{12}h^{-1}M_{\sun}$.
}
\label{tab:mass-conv-lum}
\end{deluxetable*}

\begin{figure*}
\epsscale{1.1}
\plottwo{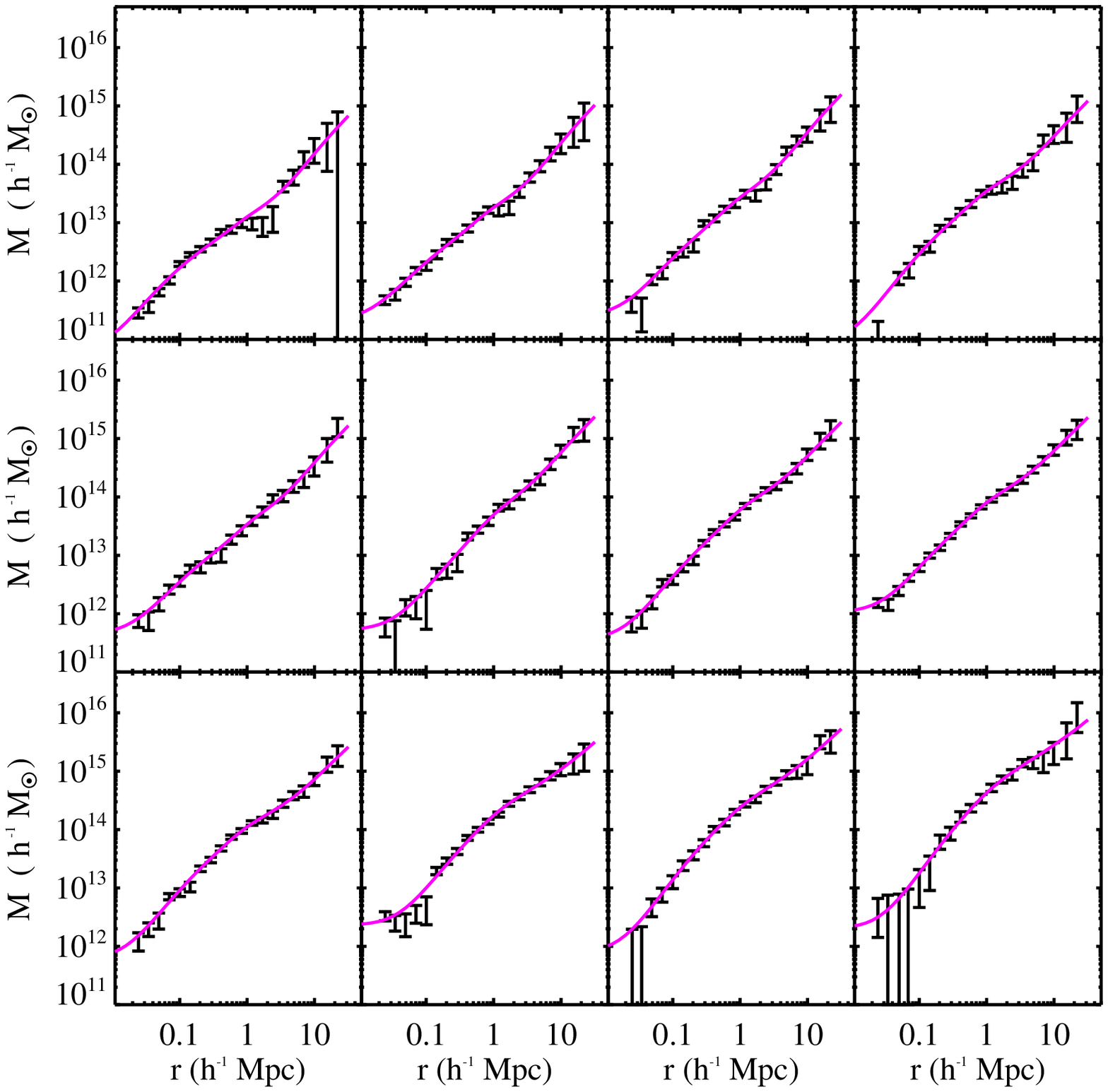}{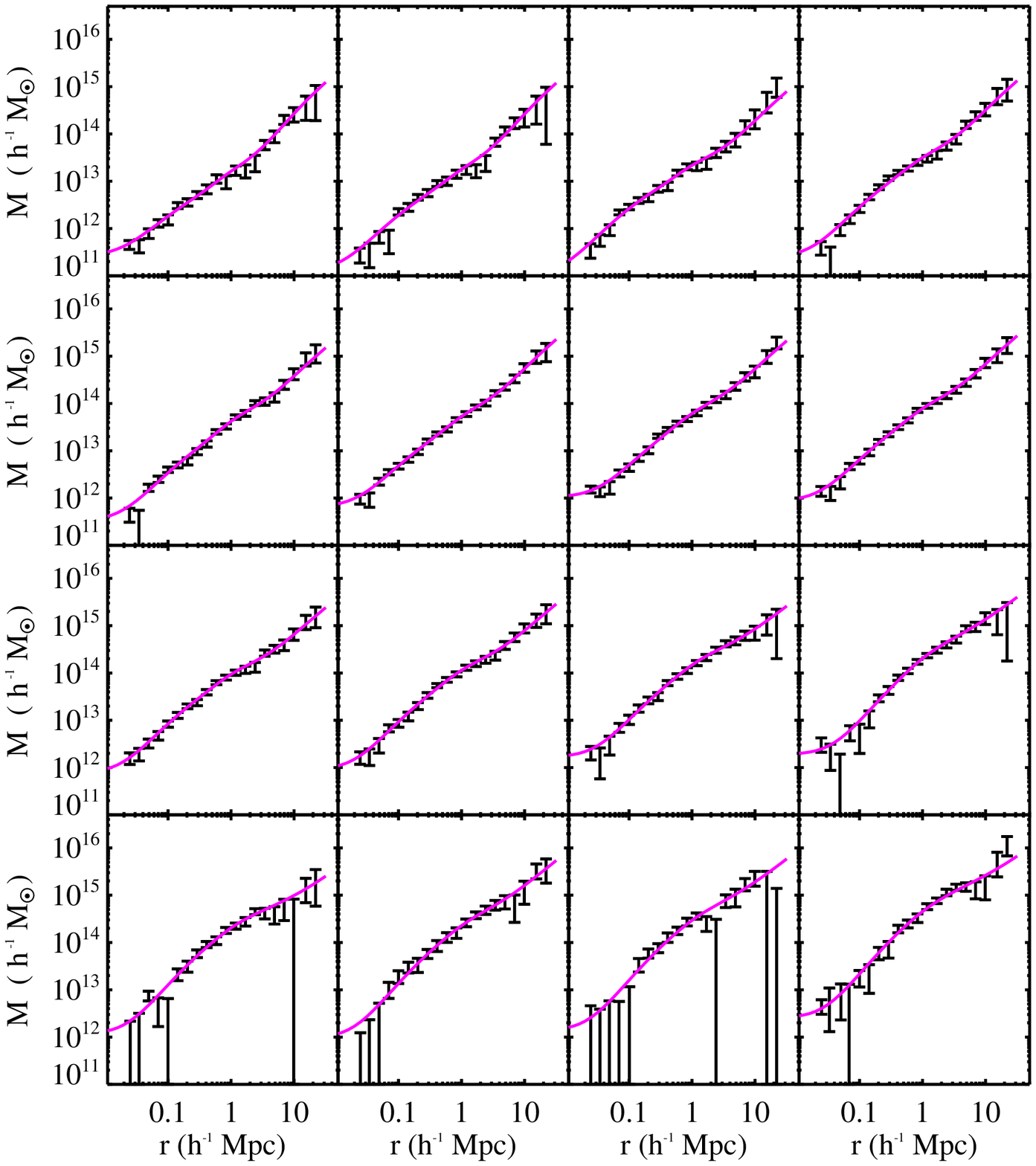}
\caption{The model fits of Fig. \ref{fig:ds-fit-n200} and  \ref{fig:ds-fit-l200} over-plotted on the inverted 3D mass profiles for
the 12 $N_{200}$ richness ({\it left panel}) and 16 $L_{200}$ luminosity bins ({\it right panel}).
\label{fig:mass-fit}}
\end{figure*}

Figure \ref{fig:mass-fit} shows the best-fit models over-plotted on the 
inverted 3D mass profiles that were previously shown in Figure \ref{fig:mass}.
%We stress that these are {\it not} fits to the inverted mass profiles themselves.
Because the mass profiles are less noisy, they more clearly 
display the features in the data. The one-halo to two-halo transition is most 
prominent in the lowest richness and luminosity bins. 

\subsection{The mass--richness relation}
\label{sec:massrich}
\begin{figure*}
\epsscale{1.1}
\plottwo{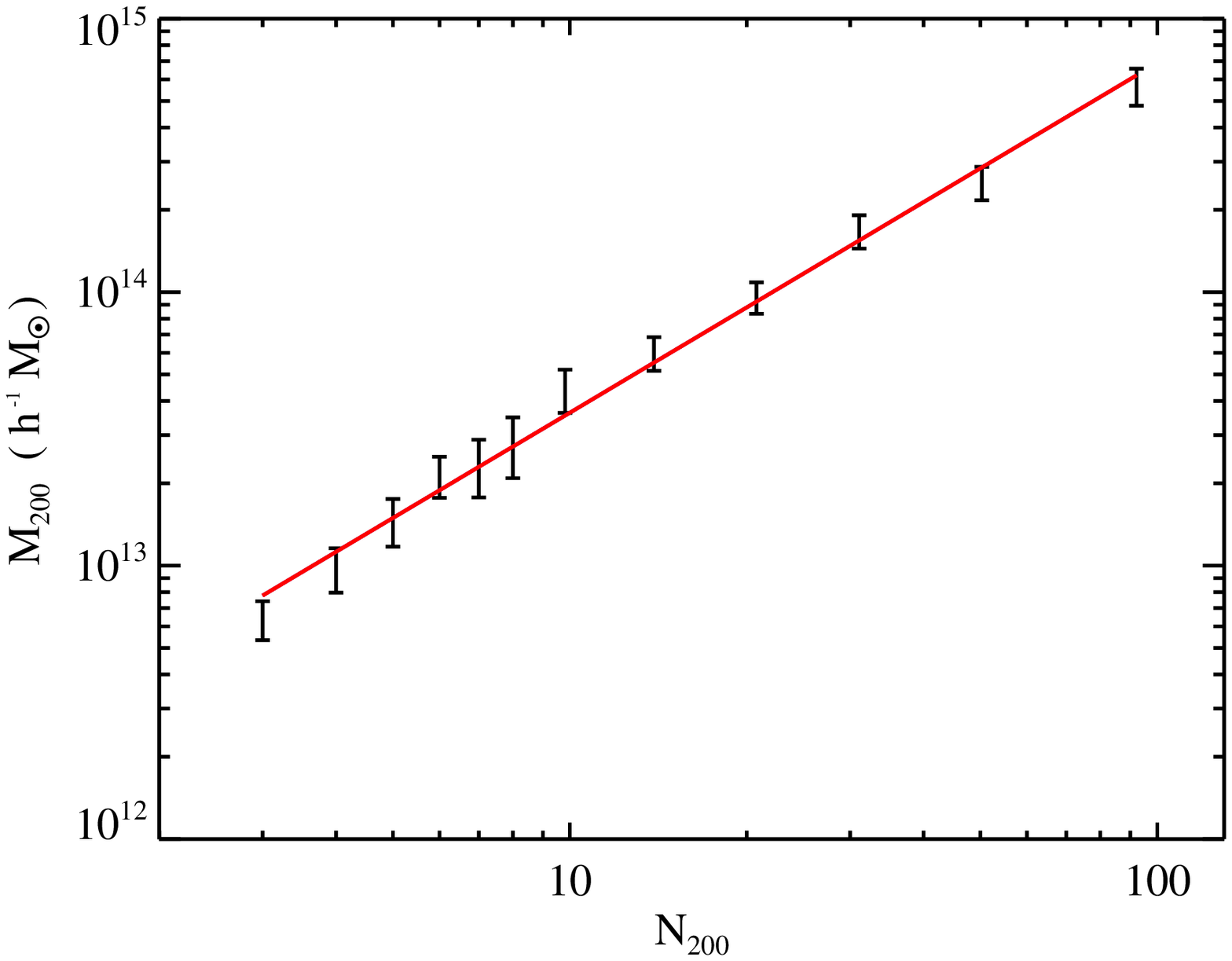}{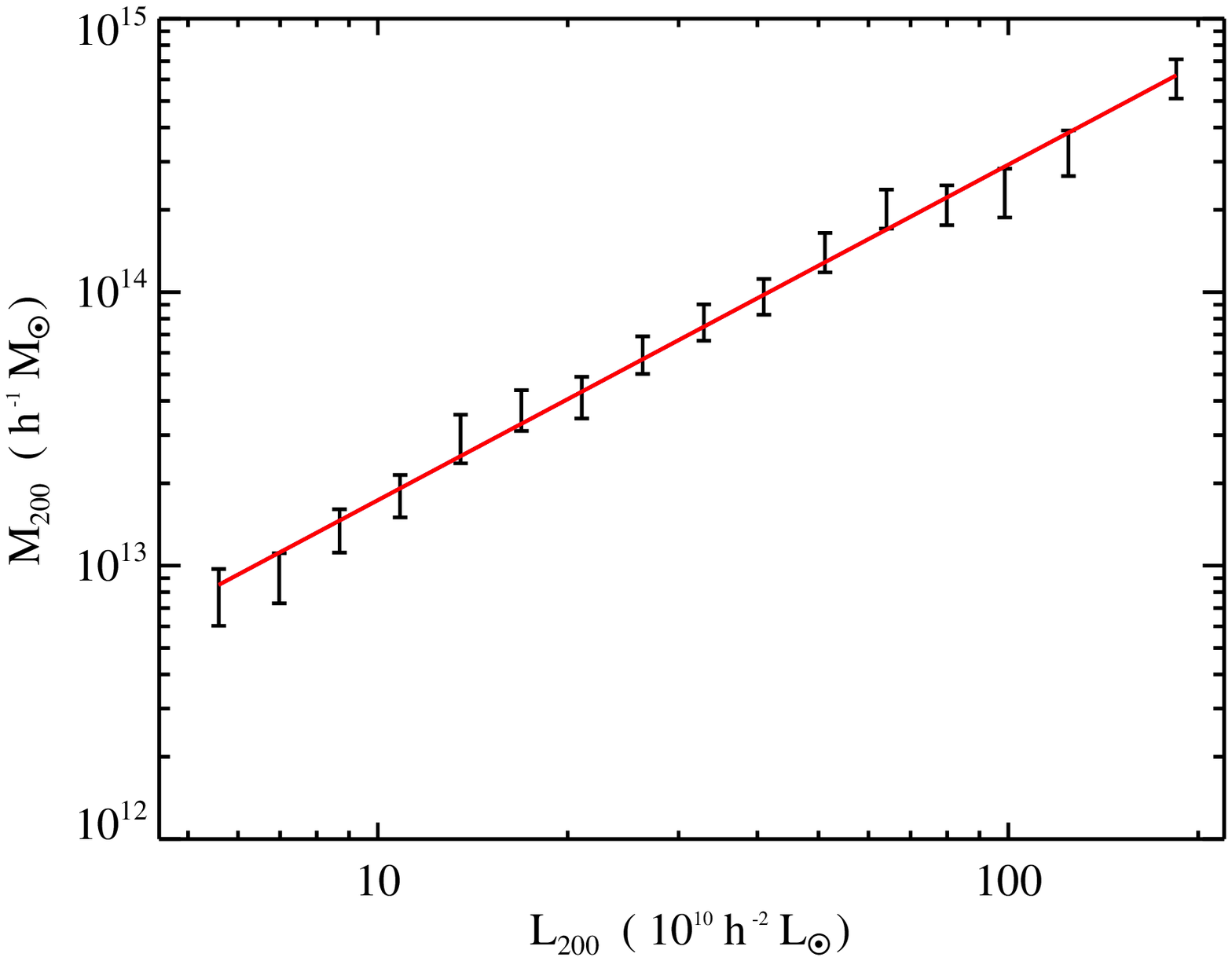}
\caption{The inferred mean halo mass vs. 
richness ({\it left panel}) and mass vs. luminosity ({it right panel}) relations from the model fits to the lensing profiles.
The red lines show the best-fit power-law relations (see text).
}
\label{fig:mass-ngal}
\end{figure*}

Figure \ref{fig:mass-ngal} shows the inferred central
halo mass--richness relations for both $N_{200}$ and $L_{200}$ richness measures.
The red line in each case shows the resulting power-law fit to the relation. The fit 
to the mass--richness relation is 
\begin{equation}
M_{200}(N_{200})= M_{200|20}(N_{200}/20)^{\alpha_N}
\end{equation}
with
\begin{eqnarray}
M_{200|20} = (8.8 \pm 0.4_{stat} \pm 1.1_{sys}) \times 10^{13} h^{-1} M_{\sun} \nonumber \\
\alpha_N = 1.28 \pm 0.04. \nonumber 
\end{eqnarray}
%\textcolor{red}{CHECK}

The mass--luminosity relation is found to be
\begin{equation}
M_{200}(L_{200})= L_{200|40}(L_{200}/40)^{\alpha_L}
\end{equation}
with
\begin{eqnarray}
M_{200|40} = (9.5 \pm 0.4_{stat} \pm 1.2_{sys}) \times 10^{13} h^{-1} M_{\sun} \nonumber \\
\alpha_L = 1.22 \pm 0.04. \nonumber
\end{eqnarray}
%\textcolor{red}{CHECK}

The statistical error on the zero-point of both mass richness relations is about 5\%. This includes the full marginalization
over the other six model parameters. As discussed in \S \ref{sec:sys}
we need to include systematic errors due to shear calibration and possible photo-$z$ biases
as well as any remaining systematic biases in our modeling.
We allow for a 3\% shear calibration bias, a 7\% photo-$z$ bias and 10\% for modeling biases.
so this increases the error on the zero-point of the mass--richness relations to about 13\%.

To accommodate other conventions used in the literature, power-law fits to the mass and concentration data for
for alternate mass-scale definitions (see Tables \ref{tab:mass-conv-ngal} and \ref{tab:mass-conv-lum}) are shown in Tables 
 \ref{tab:mass-conv-fit} and \ref{tab:mass-conv-fit-lum}.

\begin{deluxetable}{ccccc}
\tablecaption{Mass Richness Power-law Fits: $N_{200}$ Bins}
\tablewidth{0pt}
\tablehead{
\colhead{Mass type} &
\colhead{$M_{200|20}$} &
\colhead{$\alpha_N$} &
\colhead{$c_{200|20}$} &
\colhead{$\beta_N$}
}
\startdata
$M_{200}$ &    8.794E+13 &       1.28 &       3.99 &      -0.15 \\
$M_{180b}$ &    1.204E+14 &       1.30 &       6.14 &      -0.14 \\
$M_{vir}$ &    1.055E+14 &       1.29 &       5.08 &      -0.15 \\
$M_{500}$ &    6.069E+13 &       1.25 &       2.60 &      -0.16 \\
\enddata
\tablecomments{
Coefficients and exponents of the power-law fits of mass and concentration versus richness for the different virial mass definitions. The mass--richness relation and concentration--richness relation is of the form $M=M_{200|20}~(N_{200}/20)^{\alpha_{N}}$ and $c=c_{200|20}~(N_{200}/20)^{\beta_{N}}$. The \emph{relative} errors on parameters are the same as the $M_{200}$ versions (see text).
}
\label{tab:mass-conv-fit}
\end{deluxetable}

\begin{deluxetable}{ccccc}
\tablecaption{Mass Richness Power-law Fits: $L_{200}$ Bins}
\tablewidth{0pt}
\tablehead{
\colhead{Mass type} &
\colhead{$M_{200|40}$} &
\colhead{$\alpha_L$} &
\colhead{$c_{200|40}$} &
\colhead{$\beta_L$}
}
\startdata
$M_{200}$ &    9.504E+13 &       1.23 &       4.37 &      -0.15 \\
$M_{180b}$ &    1.284E+14 &       1.25 &       6.68 &      -0.14 \\
$M_{vir}$ &    1.131E+14 &       1.24 &       5.54 &      -0.14 \\
$M_{500}$ &    6.672E+13 &       1.20 &       2.86 &      -0.16 \\
\enddata
\tablecomments{
Coefficients and exponents of the power-law fits of mass and concentration versus luminosity for the different virial mass definitions. The mass--luminosity relation and concentration--luminosity relation are of the form $M=M_{200|40}~(L_{200}/40)^{\alpha_L}$ and $c=c_{200|40}~(L_{200}/40)^{\beta_L}$. The \emph{relative} errors on parameters are the same as the $M_{200}$ versions (see text).
}
\label{tab:mass-conv-fit-lum}
\end{deluxetable}

While this seven-parameter model may appear overly complicated, it is
necessary in order to properly account for the full uncertainty in
modeling the cluster shear profiles. For example, if we were to ignore
miscentering and shear non-linearity and include only the three
parameters $c_{200}, M_{200}$, and $B$ in the model fits, then the
statistical uncertainty in the calibration of the cluster
mass--richness relation would be only 3\% instead of 5\%. However, the
halo mass estimates would be biased low by a factor of $\sim
1.4$. This factor arises because $M_{200}$ is determined mostly by the
amplitude of $\Delta \Sigma$ on scales $R \lesssim 1~h^{-1}$ Mpc,
where the smoothed $\Delta \Sigma^s_{NFW}(R)$ makes very little
contribution; as a result, ignoring miscentering in fitting to the
shear on small scales leads to an underestimate of the mass by a
factor of $\sim p_c$. From the mock catalogs, we find $\langle p_c
\rangle \sim 0.7$, or $1/\langle p_c \rangle \sim 1.4$.  Therefore,
halo miscentering has a large systematic effect on the estimated halo
masses and concentrations and so must be included.

\subsection{Halo concentration scaling relations} 
\label{sec:conc} 

\begin{figure} 
\epsscale{1.2} 
\centering 
\plotone{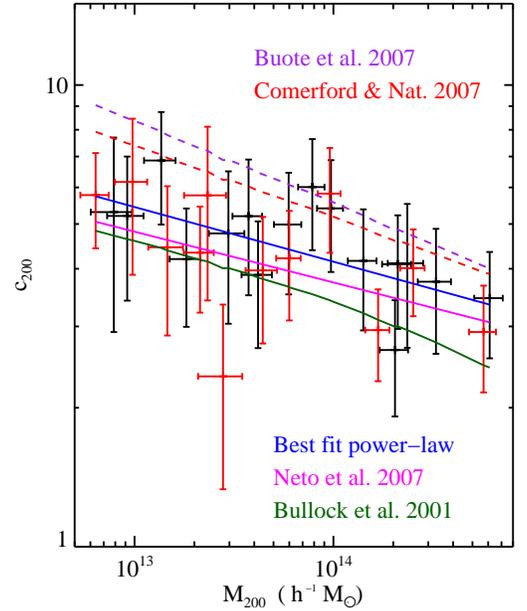} 
%\epsscale{0.9} 
%\plotone{Plots/Both_Mass_B.ps} 
\caption{
The mean NFW halo concentration parameter $c_{200}$ versus
  halo mass $M_{200}$.  Black points are from the shear profile fits
  for the $L_{200}$ luminosity bins and the red points are from the
  $N_{200}$ richness bins.  The blue curve shows the best-fit power
  law to the data (see text).  The green curve shows the prediction
  from the \citet{bullock:concentration} model with $F=0.001$,
  $K=2.9$, and our fiducial cosmology. The magenta curve shows the
  result from \cite{neto:concentration} for the Millennium Simulation
  (adjusted to $z=0.25$).  Note that this was fit to a cosmology with
  a slightly higher normalization ($\sigma_8=0.9$ vs. $\sigma_8=0.8$)
  and is thus expected to have slightly higher concentrations.  The
  purple dashed curve is a result from \cite{buote:concentration} on
  X-ray clusters; the red dashed line shows a result from a
  compilation of X-ray and strong-lensing clusters
  \citep{comerford:concentration}}
\vspace*{0.3cm} 
\label{fig:mass-c} 
\end{figure}

Figure \ref{fig:mass-c} shows the scaling of the mean concentration 
$c_{200}$ with halo mass. We have combined the results from both 
richness (red points) and luminosity bins (black points) on the same 
plot --- these are {\it not} independent, since the same clusters are 
used for both. The blue curve shows the best-fit power law, 
\begin{eqnarray} 
c_{200}(M_{200}) = c_{200|14}(M_{200}/10^{14} h^{-1}M_{\sun})^{\beta_c} \\ 
c_{200|14} = 4.1 \pm 0.2_{stat} \pm 1.2_{sys} \nonumber \\  
\beta_c = -0.12 \pm 0.04. \nonumber  
\end{eqnarray} 
%\textcolor{red}{(CHECK)}  
The fit is performed with all data points from both binnings but the
errors are adjusted upward by $\sqrt 2$ so that they are not treated
as independent data points.  These results indicate that the halo
concentrations, with typical values $c_{200} \simeq 5$, depend only
weakly on halo mass, as has been suggested by previous observational
and theoretical results.  Note that ignoring the parameters $p_c,
\sigma_s$ and $M_0$ in the model fits would lead to a (biased)
underestimate of the halo concentration parameter $c_{200}$ by about a
factor of 3, as well as to unrealistically small error estimates on
the concentration.
 
For comparison with the lensing results, the green curve in
Fig. \ref{fig:mass-c} shows the predicted concentration vs. mass
relation from the halo formation model of
\citet{bullock:concentration}. Note that \citet{bullock:concentration}
use a different definition of halo mass $M_{vir}$ and concentration
$c_{vir}$, so we have converted their predictions to our parameters
$M_{200}$ and $c_{200}$ following the translation given in the
Appendix.  In their model, the halo concentration is given by $c_{vir}
= K~(a/a_c)$, where $a=1/(1+z)$ and $a_c$ is the collapse epoch of the
halo; the time at which the typical collapsed mass, $M_*$, is a fixed
fraction $F$ of the halo mass, $M_*(a_c)=F~M_{vir}$.  This model is
defined by the two parameters $K$ and $F$, which are assumed to be
independent of cosmological parameters.  Here $M_*$ is the non-linear
mass scale at scale factor $a$ in Press-Schechter theory, i.e., the
mass for which $D(a)\sigma(M_*(a))=\delta_c$, where the linear growth
factor $D(a)$ is given by Eqn. \ref{eq:linear}, $\delta_c$=1.686 is
the critical density in the spherical collapse model, and $\sigma(M)$
is the variance of the linear density field smoothed on the scale that
on average encloses mass $M$.  We choose the parameter values $K=2.9$
and $F=0.001$ (different from the original Bullock numbers), which
have been demonstrated to reproduce the measured halo concentrations
in a more recent set of LCDM dark matter simulations
\citep{wechsler_etal:06}. With those choices, the predicted
concentrations of this galaxy formation model, shown as the green
curve in Figure \ref{fig:mass-c}, fit those inferred from the lensing
data fairly well. The $\chi^2$ between the two is 8 (for 12 degrees of
freedom) for the $N_{200}$ richness binning and 12 (for 16 degrees of
freedom) for the $L_{200}$ binning.  In making this comparison, we
have used the fiducial cosmological parameters given at the end of \S
\ref{sec:intro}. Furthermore, if we keep the Bullock $F$ parameter and
cosmological parameters fixed we can determine the best fit Bullock
$K$ parameter from our data: $K_{fit} = 3.00 \pm 0.24$ (assuming our
fiducial cosmology with $\sigma_8=0.8$).
 
Recently, \cite{neto:concentration} studied the concentrations of 
halos identified from the Millennium Simulation 
\citep{springel:millennium} and found a power-law relation for the 
average halo concentration, 
%\begin{equation} 
$c_{200} = 5.26(M_{200}/10^{14}h^{-1}M_{\sun})^{-0.1}$.   
%\end{equation} 
The Millennium simulation uses a flat LCDM cosmology with
$\Omega_m=0.25, \Omega_b=0.045, h=0.73, n_s=1, \sigma_8=0.9$ and
$z=0$.  \citet{bullock:concentration} found that halo concentration
scales as $1/(1+z)$, which is consistent with recent observational
results from X-ray clusters ; $c \propto (1+z)^{-0.71 \pm 0.52}$
\citep{schmidt_allen:concentration}.  We thus shift the
\cite{neto:concentration} relation by 0.8 to put it at our median
cluster redshift of $z=0.25$; this is shown as the magenta curve in
Fig. \ref{fig:mass-c}. This result for dissipationless halos agrees
very well with both the \citet{bullock:concentration} model and our
data ($\chi^2 = 8$).  Note that because the \cite{neto:concentration}
results are calculated for a cosmology with slightly higher
normalization ($\sigma_8=0.9$ vs. $\sigma_8=0.8$) they are expected to
have slightly higher concentrations and the agreement between the two
models is even better than it looks in the figure.  The large
difference shown in the \cite{neto:concentration} paper between their
results and the results of \citet{bullock:concentration} are due to
the fact that these authors used the original
\citet{bullock:concentration} values for $K$ and $F$, instead of the
updated ones that we use here; with this change the two theoretical
models are virtually indistinguishable, and are both in excellent
agreement with our results.
 
\cite{buote:concentration} have recently presented a determination of 
the concentration--mass relation as measured by a set of 39 clusters 
with X-ray measurements, finding $c_{\rm vir} (1+z) = (9.0 \pm 0.4) 
(M{\rm vir}/M_{14})^{-0.172 \pm 0.026}$. This is plotted as the 
purple-dashed line on Fig. \ref{fig:mass-c}. 
\cite{comerford:concentration} also recently compiled several 
concentration measurements from individual strong-lensing and X-ray 
clusters (including those of \citealt{buote:concentration} and 
\citealt{schmidt_allen:concentration}), and found $c_{\rm vir} (1+z) = 
(14.5 \pm 6.4) (M{\rm vir}/M_*)^{-0.15 \pm 0.13}$, where  $M_* = 1.23 
~10^{12} h^{-1} M_{\sun}$ at $z=0.25$ for our fiducial cosmology with 
$\sigma_8=0.8$. This is plotted as the red dashed line on 
Fig. \ref{fig:mass-c}. 

 In the figure, each of these relations is
converted to our $M_{200}$ system for comparison. Both results thus
have a mass scaling that is consistent with results; although note
that \cite{schmidt_allen:concentration} have seen some indication 
for a steeping of this power at the highest masses from a sample
of X-ray systems.

These results all have a somewhat higher normalization than our data;
there are many possibilities for this discrepancy.
%It may be that the assumptions of
%hydrostatic equilibrium or spherical symmetry are causing the
%concentration values to be overestimated for these X-ray clusters.  
At least some of the discrepancy is likely due to selection effects
between the samples.  It is likely that X-ray clusters and
strong-lensing clusters are more concentrated than average
red-sequence clusters.  In the X-ray case, they are chosen to be
relaxed systems, which likely have higher concentrations
\citep{wechsler_etal:02}.  
This effect has been estimated to
be of the order $\sim 10$--$20\%$ 
\citep{buote:concentration,schmidt_allen:concentration}
, but is still somewhat uncertain.  Very
concentrated clusters will be also more likely to produce
strong-lensing features. Also, the X-ray flux is proportional to the
square of the gas density and so X-ray selection also favors more
concentrated clusters.  It is also possible that our model for
miscentering is underestimated, which would reduce our modeled
concentrations compared to the true halo concentrations.  We may be
able to better constrain the miscentering the future, and are also
working towards measurements with a clearly well-centered cluster
sample to further investigate these effects.
%Also the clusters in \cite{buote:concentration}
%are a low-redshift sample (mean $z \approx 0.05$ versus our mean
%z=0.25) and so this difference could, in part, be due to evolution.

\cite{mandelbaum:groups}
constrains the concentration of typical halos 
containing SDSS luminous red galaxies with galaxy-galaxy lensing. They 
find $c_{180b} = 5.6 \pm 0.6$ which is $c_{200} = 3.8 \pm 0.4$,
consistent with our results. 
 
\subsection{Bias scaling relations}
\label{sec:bias}

\begin{figure}
%\epsscale{0.9}
%\plotone{Plots/Both_Mass_c.ps}
\epsscale{1.2}
\centering
\plotone{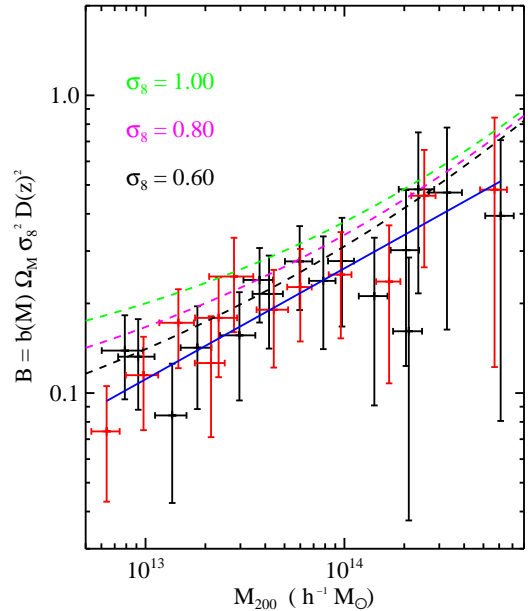}
\caption{The effective bias parameter (the coefficient of the two-halo term)  
$B = b(M_{200}) \Omega_m \sigma_8^2D^2(z)$ versus $M_{200}$: black points show 
lensing results in luminosity bins, red in $N_{200}$ bins.
The blue curve is the best-fit power law (see text). 
%$B(M_{200}) = B_{14} (M_{200}/(10^{14} h^{-1}M_{\sun})^{\beta_B}$ with
%$B_{14} = 0.33 \pm  0.02$ and $\beta_B = 0.41 \pm 0.02$.
The three dotted curves show the predictions from the 
\citet{sheth:better-bias} elliptical collapse model
for three values of $\sigma_8$: (1.0, 0.8, 0.6), from top to bottom.
}
\label{fig:mass-B}
\end{figure}

Figure \ref{fig:mass-B} shows the scaling of the mean effective bias
parameter $B$ as a function of halo mass. The lensing results are well
fit by a power law, indicated by the blue solid curve, 
\begin{eqnarray}
B(M_{200}) =B_{200|14}~(M_{200}/10^{14} h^{-1} M_{\sun})^{\alpha_B}  \\
B_{200|14} = (0.26 \pm 0.02_{stat} \pm 0.02_{sys}) \nonumber \\
\alpha_B = 0.38 \pm 0.02 \nonumber
\end{eqnarray}
%\textcolor{red}{(CHECK)}.  
The fit is performed with all data points
from both binnings but the errors are adjusted upward by $\sqrt{2}$ so
that they are not treated as independent data points.  As
theoretically expected, the clustering strength, i.e., the bias,
increases with halo mass.

As above, it is of interest to compare these results with the
predictions of structure formation models.  The halo bias can be
computed using the ``peak-background split''
\citep{mo-white:cluster-bias,sheth-tormen:cluster-bias}.  We consider
the model of \cite{sheth:better-bias}, which is derived from the
elliptical collapse model and calibrated with N-body simulations. In
their bias relation, the halo mass is defined in terms of the region
within which the mean density is 180 times the mean density of the
Universe at redshift $z$, $M_{180b} = M(r_{180b})= 4/3 \pi r_{180b}^3
180\bar{\rho}(z)$.  Using the formulas of the Appendix, we convert between this
definition and our expression for $M_{200}$ given in
Eqn. \ref{eq:virMr}.  The \citet{sheth:better-bias} halo bias relation
is given by

\begin{eqnarray}
& & b(\nu) = 1 + \frac{1}{\sqrt{a} \delta_c} \times \nonumber \\
& & \left[ \sqrt{a}(a \nu^2) + b\sqrt{a}(a \nu^2)^{1-c} 
-\frac{(a \nu^2)^c}{(a \nu^2)^c + b (1-c)(1-c/2)} \right] \nonumber \\
\label{eq:smtbias}
\end{eqnarray}

\noindent where $\delta_c=1.686$ and
$\nu=\delta_c/(D(z)\sigma(M))$. \citet{sheth:better-bias} chose
parameters $a=0.707, b=0.5$, and $c=0.6$ to agree with N-body
simulations. However, both \cite{seljak-warren:halo-bias} and
\cite{tinker:ml} determined that this relation over-estimates the bias
at fixed halo mass by about $20\%$, especially for masses less than
the non-linear mass scale $M_*$.  \citet{tinker:ml} find that this
expression gives a better fit to the simulations for $a=0.707,
b=0.35$, and $c=0.8$, and we adopt these parameter values to compare
with the lensing results.

One effect that needs to be included is that we are not measuring $B(M_{200})$ exactly
but rather $\left<B(M_{200})\right>$ where the average is over the log-normal distribution
of mass. Similarly, we are plotting these versus $\left<M_{200}\right>$. Therefore, to 
compare the theoretical predictions to the data we need to multiply the theoretical
predictions at $\left<M_{200}\right>$ by $\left<B\right>/B(\left<M_{200}\right>)$ which is
$\exp(V_M ~\alpha_B (\alpha_B-1)/2)$ for a log-normal distribution. Here, $\alpha_B = 0.38$, is the
logarithmic slope $B(M_{200}) \sim M_{200}^{\alpha_B}$ and $V_M$ is the variance 
of $\ln M_{200}$. This correction varies with richness
but is typically about 10\% and adds about $5-10\%$ uncertainty to the predictions
depending on the width of the prior distribution for $V_M$. With our (probably overly generous) 
prior of $0.6$ for $\ln V_M$, this uncertainty is 10\%.

The resulting theoretical expressions for $B$ are plotted as the dashed lines
(black, magenta, and green) in Fig.  \ref{fig:mass-B} for three
values of $\sigma_8$: $0.6, 0.8$, and $1.0$. These correspond to non-linear masses, $M_*$:
$0.43,1.23,$ and $5.26 \times 10^{12} h^{-1} M_{\sun}$ at $z=0.25$.
Although the predictions for all three choices are within $\sim 30\%$ of the best-fit relation
from lensing, the data appear to prefer lower values of $\sigma_8$. The $\chi^2$s are
acceptable for both $\sigma_8=0.6$ ($\chi^2_{N}=7, \chi^2_{L}=12$)
and $\sigma_8=0.8$ ($\chi^2_{N}=15, \chi^2_{L}=20$) 
but formally unacceptable for $\sigma_8=1$ ($\chi^2_{N}=32, \chi^2_{L}=36$). The
number of degrees of freedom is 12 and 16 for $\chi^2_{N}$ and $\chi^2_{L}$ respectively.
These $\chi^2$ numbers do \emph{not} include the above mentioned $V_M$ uncertainty and
so can be reduced by another 10-20\%.
We refrain from drawing cosmological conclusions from this comparison for
several reasons. First, in fitting the halo model to the lensing
results, we assumed particular values for the cosmological parameters (except $\sigma_8$)
when we calculated the linear correlation
function $\xi_l$ (Eqn. \ref{eq:r2h2}) for the two-halo term. For a self-consistent
cosmological constraint, we would need to float the cosmological
parameters in calculating the two-halo term for the lens model fit. It
would also be desirable to allow for possible scale-dependent bias,
since the predicted non-linear correlation function at the largest
scales we probe, $25-40 ~h^{-1}$ Mpc, differs slightly from the linear
theory prediction \citep{smith:bias}. We would also want to consider halo-exclusion effects
\citep{zheng:halo-excl}. We believe that precise prediction of the bias involving all of these
effects at these intermediate scales is not yet possible but clearly the quality of data
is improving to the point where such study is now warranted.

It would be better to extend the lensing measurements to slightly larger scales ($\ge 50 ~h^{-1}$
Mpc comoving) in order to reduce this effect and, more importantly, to
isolate the large-scale bias measurements from degeneracies with the
NFW halo parameters. Finally, to reliably estimate cosmological
parameters we would require data with better signal-to-noise ratios as well
as more precise shear and photometric redshift calibration. In future
wide-area, deeper lensing surveys, these conditions will all be met, and
constraints on cosmology from lensing measurements of the halo bias
will be possible.

\subsection{BCG--halo mass scaling relation}

\begin{figure}
\epsscale{1.1}
\plotone{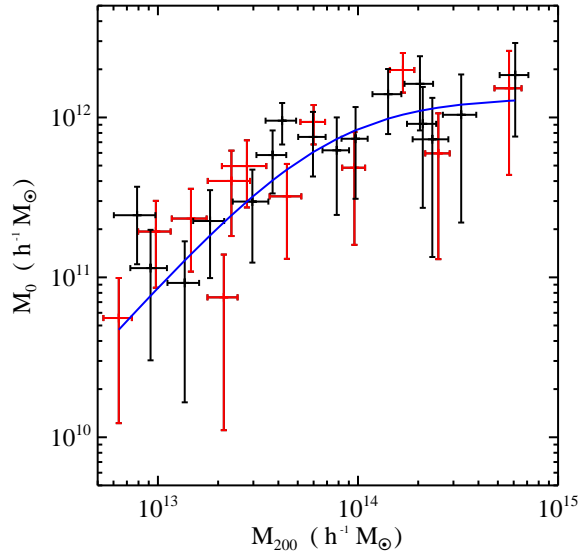}
\caption{
The BCG point mass vs. mean central halo mass for both richness bins: $N_{200}$ (red) $L_{200}$ (black).
The point mass term increases with central halo mass at low halo mass but flattens out at 
about $10^{12} h^{-1} M_{\sun}$. The blue curve is a fitting function: $M_0 = p_0/(1+(M_{200}/p_1)^{p_2})$
with $p_0=1.334~10^{12} h^{-1} M_{\sun}$, $p_1=6.717~10^{13} h^{-1} M_{\sun}$ and $p_2=-1.380$. The point mass
is roughly consistent with the masses of luminous red galaxies from strong-lensing constraints. 
}
\label{fig:point-mass}
\end{figure}

We have included this point mass term in our model mostly to allow enough freedom so that the 
concentration measurements would not be overly influenced by the first few data points.
This is especially important when non-linear shear is considered.
However, the relation between the BCG mass and the central halo mass may be of interest in itself.
Figure \ref{fig:point-mass} shows the point mass term, $M_0$, plotted versus the mean central halo mass, $M_{200}$.
The point mass increases with central halo mass but seems to plateau at an asymptotic mass of about $1.3 \times 10^{12} h^{-1} M_{\sun}$.
The blue curve is simply a fitting function: $M_0 = p_0/(1+(M_{200}/p_1)^{p_2})$
with best fit values $p_0=1.334 \times 10^{12} h^{-1} M_{\sun}$, $p_1=6.717 \times 10^{13} h^{-1} M_{\sun}$ and $p_2=-1.380$.
These masses are consistent with the expected masses of galaxy halos.
Strong-lensing constrains (e.g. \cite{rusin:sis-lens}) show that nearly every strong lens is well fit by an
singular isothermal sphere out to at least $100 h^{-1}$ kpc. The 3D mass of a singular isothermal sphere is given by
$M_{SIS}(r) = 4.64 \times 10^{12}~h^{-1} M_{\sun}~(\sigma_s/100~\mbox{km/s})^2 ~(r/\mbox{Mpc})$ which at $25 h^{-1}$ kpc gives
$1.2 \times 10^{11}, 4.6 \times 10^{11}$ and 
$1 \times 10^{12} h^{-1} M_{\sun}$ for stellar velocity dispersions, $\sigma_s = $100, 200 and 300 km/s respectively.
This mass range agrees well with our point-mass values. This comparison is inexact since SISs and point
masses have different shear profiles. A precise measurement of the mass density of the central BCG would be better
suited to a combination of strong and weak lensing (e.g. \cite{gavazzi:slacs}) and is beyond the scope of this paper.

\section{Comparison of lensing and dynamical mass measurements}
\label{sec:dynamical}

\begin{figure}
\epsscale{1.1}
\plotone{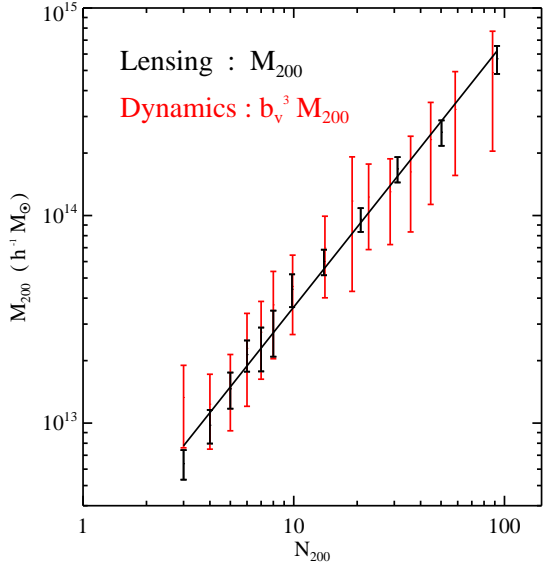}
\caption{Mean halo mass vs. richness from the lensing profiles (black points, the 
same as those shown in Fig. \ref{fig:mass-ngal}) and from 
dynamical galaxy velocity dispersion measurements (red points) from the same cluster sample 
\citep{becker:vel}. The two scaling relations are in good agreement, 
although the lensing data provides a tighter relation. Black curve shows the 
best-fit power-law relation from the lensing data. 
}
\label{fig:lensing-dynamics}
\end{figure}

\citet{becker:vel} have recently estimated statistical masses of MaxBCG
clusters from the \citet{koester:maxbcg-cat} catalog from
stacked velocity dispersion measurements.  Using galaxies near
each BCG with measured spectroscopic redshifts, they build a richness-dependent histograms 
of velocity differences and fit the shape to a summed, log-normally  distributed, set of 
Gaussians.  Results show that the geometric mean velocity dispersion scales as a power law, 
$\sigma_v \sim N_{200}^{0.436 \pm 0.015}$, with a log-normal dispersion that declines 
from $0.40 \pm 0.02$ at $N_{200}=10$ to $0.15 \pm 9$ at $N_{200}=88$.   Although the 
typical maxBCG cluster contains few galaxies with spectroscopic redshifts, it is the 
case that, as with cross-correlation lensing, the velocity histograms can be stacked 
from many clusters within a richness bin to build a high signal-to-noise histogram of
the average velocity differences. 
The best-fit velocity dispersion implies a mass, $M_{200}$, derived from the dark matter 
virial relation, $\sigma_{\rm DM}(M_{200},z) = (1082.9 \pm 4.0~{\rm km/s})
(h(z)M_{200}/10^{15} M_{\sun})^{0.3361 \pm 0.0026}$, calibrated recently from a suite of N-body 
simulations \citep{evrard:mass-vel}.  Galaxy and dark matter dynamics may differ, and this 
potentiality is approximately treated as a constant velocity bias parameter, $b_v \equiv \sigma_v / \sigma_{\rm DM}$.  
Figure \ref{fig:lensing-dynamics} shows the mean virial mass estimates from the
dynamical measurements in red 
%\textcolor{red}{Should use different symbols in addition to color here.  
%Change ``Dynamics'' label to $b_v^3 M_{\rm vir}$ or similar.} 
and the lensing halo masses (from Fig. \ref{fig:mass-ngal})
in black.  The line shows the best-fit, power-law relation from
the lensing masses.  The \citet{becker:vel} error bars
include systematic errors inherent in their method, and are
correlated. 
The statistical lensing and dynamical mass estimates appear in very good agreement, but systematic 
uncertainties in $b_{v}$ remain to be understood.  When
\citet{becker:vel} tested their virial mass estimator on mock SDSS maxBCG catalogs, they found it 
to systematically underestimate halo masses by 25\%.  This correction factor has not been applied 
to the estimates in Fig. \ref{fig:lensing-dynamics}. Including it would elevate $b_v^3 M_{\rm vir}$ 
above the lensing masses, suggesting a positive velocity bias, $b_v \simeq 1.1$.  The current level 
of agreement indicates that the velocity bias parameter is not significantly different from unity.  
We defer a more formal analysis of these issues to future work.

\section{Discussion}

In Paper I of this series, we demonstrated that cross-correlation weak
lensing can be measured around clusters of galaxies out to large
radii, $R \sim 30 ~h^{-1}$ Mpc. In this work, we have shown that these
mean shear profiles are well described by realistic models derived
from N-body simulations. Our primary results are the lensing
calibration of the cluster halo mass--richness and mass--luminosity
relations and measurements of the scaling relations between the mass,
bias, and concentration of halos. 
We also show that lensing-inferred masses are consistent with
estimates from stacked velocity dispersion measurements
\citep{becker:vel} as long as the velocity bias parameter is not significantly different from unity. 
The scaling relation between
halo concentration and mass that we derive from lensing agrees well
with the results of N-body simulations (e.g., the model of
\citealt{bullock:concentration} as updated by \citealt{wechsler_etal:06},
or the recent results of \citealt{neto:concentration}).  The scaling
between halo bias and mass from lensing is in agreement with the
simulation-calibrated predictions of \citep{sheth:better-bias}.

In this work, we have limited the analysis to the modeling of the
lensing profiles. However, for completeness we now describe some
cosmological applications of these results that are now possible. We
then conclude by suggesting some applications of these methods that
will be possible with the ambitious wide-field surveys now being
planned.

Perhaps the most obvious application of the measured halo
mass--richness relation is to measurement of the mass function of
clusters. Previously, \cite{rozo:mass-function1} completed a first
analysis constraining cosmology through the cluster mass function
using the same SDSS cluster catalog. They modeled the mass--richness
relation using the Halo Occupation Distribution (HOD) model, without a
strong observational prior on the mass--richness relation itself.  In
this model, one adopts a parametrized mass--richness relation, and
\citet{rozo:mass-function1} employed tight cosmic microwave background
and SN Ia priors on cosmological parameters \emph{except} for
$\sigma_8$, for which only a non-informative prior was used. They
found $\sigma_8 = 0.92 \pm 0.1$ and derived constraints on the HOD
model parameters. This method \citep{rozo:methods} employs
marginalization over a generous supply of nuisance parameters that
connect the observables to mock catalog predictions (Wechsler et
al. in preparation).  While \citet{rozo:mass-function1} represents one of
the more robust measurements of $\sigma_8$ from the cluster mass
function, an update to this work using the mass--richness relation
derived from lensing is in progress. This should allow for tighter
constraints on $\sigma_8$, a tight constraint on $\Omega_m$, as well
as a more precise measurement of the HOD parameters.

\cite{mandelbaum:robust-s8} have put a lower bound on $\sigma_8
(\Omega_m/0.25)^{0.5} > 0.62 $ at 95\% C.L. by employing a method
simpler than full modeling of cluster number counts. They argue that
the lensing signal around a sample of isolated luminous red galaxies
in the SDSS could not be produced by low values of $\sigma_8$ since too
few clusters would have formed. Interpretational complications such as incompleteness of
the cluster sample or miscentering would only decrease the predicted signal, 
so their bound should be robust. 

There are several ways in which measurements of stacked lensing
profiles around clusters can be used to derive entirely new constraints on
cosmology. The amplitude of the linear galaxy or cluster 
auto-correlation function measures the combination
$b^2\sigma_8^2D^2(z)$, whereas lensing measures
$b\sigma_8^2D^2(z)\Omega_m$.  Combining both galaxy or cluster
auto-correlations with lensing will thus allow one to measure the two
combinations $\Omega_m\sigma_8D(z)$ and $b/\Omega_m$.
By combining these two measurements into an estimate of 
$\Omega_m\sigma_8D(z)$, it is possible to directly probe the growth
of structure. The linear growth factor is sensitive to cosmological
parameters affecting the Hubble parameter, such as $\Omega_m$, as well
as to dark energy and spatial curvature. This growth measurement 
would complement geometric
probes of dark energy such as type Ia  supernovae and baryon acoustic oscillations.
In addition, this measurement of the growth factor would complement
cluster number counts since it extracts information from much larger
scales.  Measuring dark energy through this direct measurement of the
growth factor would have very different systematics from both cosmic
shear and cluster number counts.  This measurement will most likely require
lensing measurements extending to slightly larger scales, $50-100 h^{-1}$ Mpc,
to better tie down the $B$ parameter, as well as more attention to
systematic errors such as shear calibration and photo-$z$
biases. Since it relies on large-scale information, it will require a deep survey over
a large fraction of the sky to reduce the cosmic variance to a
small enough level to compete with the other methods.

\cite{seljak:bias} employed a similar technique to constrain
$\sigma_8$ by using lensing to constrain halo masses so that halo
biases could be predicted and used to ``de-bias'' the galaxy power
spectrum. This however requires the complication of HOD modeling to
connect galaxies to the halos that they occupy. It could be
simpler to apply this idea directly to clusters, since this requires only
large-scale auto-correlation function (or power spectrum) measurements and 
does not require large-scale lensing measurements.  This approach does,
however, rely on models for the bias prediction
\citep[e.g.][]{sheth-tormen:cluster-bias,seljak-warren:halo-bias}, 
so direct measurement of the bias would be preferred
as long as the errors are sufficiently small.

Future weak lensing surveys such as SNAP \citep{snap:de}, DUNE
\citep{refregier:dune}, LSST \citep{tyson:lsst} and DES \citep{des:05}.
would be ideal for these types of measurements.  The statistical errors on the average
shear in a radial bin should be at the percent level for these surveys, compared
to ~50\% for the SDSS cluster data for identical binning.
Since the dark energy constraints from measurement of the
shear power spectrum will already require shear calibration and photo-$z$
biases below the percent level, this would suggest that these surveys
should be able to measure halo masses, concentrations, and biases at about the percent
level for perhaps hundreds of richness bins. Entirely new ways of
using lensing to constrain cosmology may be possible. For example,
baryon acoustic oscillations should leave their imprint on the
$\Delta\Sigma$ profile at comoving scales of $100 h^{-1}$ Mpc and will
be detectable with surveys such as these.  Determining how to extract
the most information from such a data set should remain a fruitful
area of study.

\acknowledgments
The research described in this paper was performed in part at the Jet
Propulsion Laboratory, California Institute of Technology, under a
contract with the National Aeronautics and Space Administration.  ESS
was supported by NSF grant AST-0428465.  BK, TAM, AEE and MRB
gratefully acknowledge support from NSF grant AST 044327 and the
Michigan Center for Theoretical Physics.  RHW received partial support
from the U.S. Department of Energy under contract number
DE-AC02-76SF00515.  ER was funded by the Center for Cosmology and
Astro-Particle Physics at The Ohio State University. JF acknowledges 
support from the DOE at Fermilab and the University of Chicago and 
from the Kavli Institute for Cosmological Physics at Chicago.

\appendix
%{Conversion between different mass definitions}

``Virial-type''  mass definitions all have the form

\begin{equation}
M_a = M(r_a) = \frac{4 \pi}{3} r_a^3~\Delta_a~\rho_a~,
\end{equation}

\noindent where $M(r)$ is the mass profile. The number
$\Delta_a$ may be a function of cosmology and redshift. Typical
choices for $\Delta_a$ are 200 and 180.  The density $\rho_a$ is
always some variation of the critical density, $\rho_{crit}$, but it may
be $\rho_{crit}(z)$ or $\bar{\rho}(z) = \rho_{crit}(z=0)~(1+z)^3$. Let
us define $D_a \equiv \Delta_a~\rho_a$, since the conversion between 
different conventions only depends on this product.

For any two choices of $D_a$, there is a conversion between them for the
mass $M_a$, (or equivalently $r_a$) and the NFW concentration parameter
$c_a$.  $M_a$ and $c_a$ (unlike $r_s$ and $r_a$) are independent of the
choice between \emph{physical} and \emph{comoving} units.

\cite{hu-kravtsov:sample-var} discuss this issue but we will
review the conversion again here.

The NFW form for the density profile is given by

\begin{equation}
\rho(r) = \frac{\rho_s}{(r/r_s)~(1+r/r_s)^2}~.
\end{equation}

\noindent Under this assumption, the mass profile for some choice of mass definition is given by

\begin{equation}
M(r) = 4 \pi \rho_s r_a^3 ~f(r_s/r_a)~,
\end{equation}

\noindent where
 
\begin{equation}
f(x) = x^3 \left[ \ln(1+x^{-1}) - (1+x)^{-1} \right]~,
\end{equation}

\noindent and the concentration is defined as $c_a= r_a/r_s$.  The parameters
$r_s$ and $\rho_s$ are independent of the choice of $D_a$,  so for any
other choice $D_b$ we have $3 \rho_s=D_a/f(1/c_a) = D_b/f(1/c_b)$. 
Therefore we have the conversion between the two concentrations, 

\begin{equation}
1/c_b = f^{-1}\left(\frac{D_b}{D_a} ~f(1/c_a)\right)~.
\end{equation}

\noindent Similarly, $r_s = r_a/c_a = r_b/c_b$, so the conversion for these
``virial'' radii is $r_b= r_ac_b/c_a$, and the conversion between
masses is

\begin{equation}
M_b = M_a ~\frac{D_b}{D_a}~ \left(\frac{c_b}{c_a}\right)^3~.
\end{equation}

\noindent The inverse function of $f$ needs to be computed with a look-up table
and interpolation since a simple closed-form expression does not
exist. However, the conversion simply depends on the ratio $D_b/D_a$.

An as example, we consider the two most common choices, $D_{200c} =
200\rho_{crit}(z) = 200\rho_{crit}(0)H^2(z)/H_0^2=
200\rho_{crit}(0)[\Omega_m(1+z)^3 + (1-\Omega_m)] $ (in a flat LCDM
universe) and $D_{180b} = 180\bar{\rho}(z) =
180\rho_{crit}(0)(1+z)^3~\Omega_m$. The ratio of these is

\begin{equation}
\frac{D_{180b}}{D_{200c}} = \frac{9}{10}~\Omega_m(z) =
\frac{9}{10}~\frac{\Omega_m~(1+z)^3}{\Omega_m~(1+z)^3 + (1-\Omega_m)}
\end{equation}

We use this formula to convert our measured masses
$M_{200c}$ to $M_{180b}$, using $z=0.25$ and $\Omega_m =0.27$, which
gives $D_{180b}/D_{200c} = 0.377$. We use this conversion to compute the halo bias, since 
it has been shown to be nearly universal when expressed in the
$D_{180b}$ mass definition.

Similarly, to calculate the halo concentration using the \citet{bullock:concentration} 
model, we
need to convert $M_{200}$ to $M_{vir}$. This conversion uses
\citep{bryan-norman:collapse}

\begin{equation} 
\Delta_{vir} \equiv \frac{18 \pi^2 + 82 x -39 x^2}{1+x}~,
\end{equation}

\noindent with $x \equiv \Omega_m(z)-1$. This results in 

\begin{equation}
\frac{D_{vir}}{D_{200c}} \equiv \frac{18 \pi^2 + 82 x -39 x^2}{200}~.
\end{equation}

%\begin{singlespace}
%\bibliographystyle{apj}
%\bibliography{biblio}

%\end{singlespace}

\end{document}